\documentclass[leqno]{article}

\usepackage{graphicx}
\usepackage{color}
\usepackage{array}
\usepackage{fullpage}
\usepackage{fancyhdr}
\usepackage{amsmath,amsthm}
\usepackage{amssymb,latexsym}
\usepackage{mathrsfs}
\usepackage{cancel}
\usepackage[initials,short-journals,short-months]{amsrefs}
\usepackage[pdfstartview={FitH},colorlinks]{hyperref}
\usepackage{url}
\usepackage{tikz}
\usepackage{enumitem}
\usepackage{soul}
\usepackage{bbm}
\usepackage[FIGTOPCAP]{subfigure}

\theoremstyle{definition}

\theoremstyle{remark}

\def\XXint#1#2#3{{
\setbox0=\hbox{$#1{#2#3}{\int}$}
\vcenter{\hbox{$#2#3$}}\kern-.5\wd0}}

\def \CSmooth(#1,#2){\mathcal{C}_{#1,#2}}

\def \Mgale(#1,#2){M_{#1}^{#2}}
\def \Ngale(#1,#2){\mathcal{N}_{#1}^{#2}}

\title{Heat kernels on 2d Liouville quantum gravity: a numerical study\thanks{Preliminary report. Research partially supported by NSF DMS-1106982.}}
\author{
Grigory Bonik
\\ \small{University of Connecticut}
\and
Joe P. Chen
\\ \small{University of Connecticut}
\and
Alexander Teplyaev
\\ \small{University of Connecticut}
}

\date{\today}

\numberwithin{equation}{section}

\begin{document}

\maketitle

\begin{abstract}
We numerically compute the heat kernel on a square lattice torus equipped with the measure corresponding to Liouville quantum gravity (LQG). From the on-diagonal heat kernel we verify that the spectral dimension of LQG is $2$. Furthermore, when diffusion is started from a high point of the underlying Gaussian free field, our numerics indicates superdiffusive space-time scaling with respect to the Euclidean metric in the small space-to-time regime. The implications of this result require further investigation, but seem to coincide with the notion that the Euclidean metric is \emph{not} the right geodesic for characterizing the geometry of LQG. 
\end{abstract}

\section{Introduction}

Liouville quantum gravity (LQG)  is a 2-dimensional space endowed with a random metric $g$ of the form
\begin{equation}
g(z) = e^{\gamma X(z)} \,dz^2,
\end{equation}
where $X$ is a random ``matter field'' on the space, and $\gamma$ is a coupling constant which depends on the ``central charge'' of the matter field. For certain values of the central charge, one can show that $X$ becomes the Gaussian free field. (For more details about the free field the reader is referred to the survey \cite{SheffGFF}.) Giving rigorous sense to this random metric $g$ is a mathematical challenge which has sparked much interest recently. To address this problem, Duplantier and Sheffield \cite{DupShe} introduced a random measure, henceforth called the Liouville measure, on a two-dimensional domain $\mathbb{D}$ of the form
\begin{equation}
M_\gamma = e^{\gamma X(z)-\frac{1}{2}\gamma^2\mathbb{E}[X^2(z)]} \,dz^2,
\end{equation}
and established the scaling relations of the dimension of typical sets in the Euclidean metric vs. in the quantum metric, known as the Knizhnik-Polyakov-Zamolodchikov (KPZ) relation in the physics literature. Later, Garban, Rhodes, and Vargas \cite{GarbanRhodesVargasI} and Berestycki \cite{BeresLBM} constructed a Brownian motion on $(\mathbb{D}, M_\gamma)$ as a time-change of ordinary 2-dimensional Brownian motion. In \cite{GarbanRhodesVargasII} the authors further established the Feller property of Liouville Brownian motion and characterized the associated Dirichlet form. More recently, the authors of \cite{MRVZ} and \cite{AndresKajino}, respectively, proved estimates of the heat kernel associated with Liouville Brownian motion on the torus and on the whole plane.

The goal of our project is to simulate LQG on a square lattice and numerically solve the heat equation in order to obtain more detailed information about the heat kernel. Let $\mathbb{T}_n = (\mathbb{Z}/n\mathbb{Z})^2$ be the square torus of side $n$, taken to be an integer power of $2$. Let us equip $\mathbb{T}_n$ with the random measure
\begin{equation}
\label{eq:cov}
 \quad M_{n,\gamma}(dx, \omega)= \exp\left(\gamma X_n(x,\omega) - \frac{\gamma^2}{2}\mathbb{E}[X_n^2(x)] \right) dx,
\end{equation}
where $dx$ is the counting measure on $\mathbb{T}_n$ normalized by $n^{-2}$, $\gamma \in [0,2)$ is a parameter indexing the family of subcritical LQG models, and $\{X_n(\omega,x)\}_{x\in \mathbb{T}_n}$ is a realization of the discrete free field on $\mathbb{T}_n$, which is a family of centered Gaussian random variables with covariance
\begin{equation}
\mathbb{E}[X_n(x) X_n(y) ] = (2\pi) G_{\mathbb{T}_n}(x,y), \quad\forall x,y\in \mathbb{T}_n.
\end{equation}
where $G_{\mathbb{T}_n}: \mathbb{T}_n \times \mathbb{T}_n \to \mathbb{R}_+$ is the Green's function associated with the Laplacian on $\mathbb{T}_n$. The factor $2\pi$ is important for normalization purposes.

In a nutshell, we are changing from the uniform counting measure to a new measure whose Radon-Nikodym derivative is a constant multiple of $e^{\gamma X_n}$. However, since the maximum of the free field $X_n$ a.s. diverges like $2\log n$ \cite{BDZ11}, the measure $M_{n,\gamma}$ becomes singular with respect to the uniform measure in the $n\to\infty$ limit. The exponential $\exp\left(-\frac{\gamma^2}{2} \mathbb{E}[X_n^2(x)]\right)$ is a renormalization factor which makes $(M_{n,\gamma})_n$ into a martingale along an approximating sequence $(X_n)_n$ of the continuous free field. From now on we will fix a value of $\gamma \in (0,2)$ and suppress the index.

The Dirichlet form for ``Liouville random walk'' on $\mathbb{T}_n$ is
\begin{equation}
\label{eq:LQGDF}
\mathcal{E}_n(f,g) = \frac{1}{2} \sum_{x,y\in \mathbb{T}_n} [f(x)- f(y)][g(x)-g(y)]\mu_{xy},
\end{equation}
where $\mu_{xy} = 1/4$ if and only if $x$ and $y$ are connected by an edge in $\mathbb{T}_n$. This looks just like the classical Dirichlet form for symmetric random walk on $\mathbb{T}_n$. However, for LQG, the process generated by (\ref{eq:LQGDF}) is symmetrized by the measure $M_n$, not by $dx$. Thus to write down the infinitesimal generator $\Delta_{M_n}$ we need to use the identity
\begin{equation}
\mathcal{E}_n(f,g) = \langle f, -\Delta_{M_n} g \rangle_{L^2(\mathbb{T}_n, M_n)}.
\end{equation}
It is straightforward to verify that
\begin{equation}
\label{eq:LQGLap}
\Delta_{M_n}(\omega) = e^{-\gamma X_n(\omega) + \frac{\gamma^2}{2} \mathbb{E}[X_n^2]}\Delta,
\end{equation}
where $\Delta$ is the graph Laplacian on $\mathbb{T}_n$. 

Note that by (\ref{eq:cov}), $\mathbb{E}[X_n^2(x)]$ is equal to $(2\pi)$ times the on-diagonal Green's function $G_{\mathbb{T}_n}(x,x)$ on the torus $\mathbb{T}_n$. Since all points are isotropic on the torus,
$$
G_{\mathbb{T}_n}(x,x) = \frac{1}{n^2}\sum_{x\in \Lambda_n} G_{\mathbb{T}_n}(x,x) = \frac{1}{n^2} {\rm Tr} \,G_{\mathbb{T}_n} = \frac{1}{n^2} \sum_j \frac{1}{\lambda_j},
$$
where $\{\lambda_j\}_j$ denotes the eigenvalues of the (ordinary) Laplacian on $\mathbb{T}_n$. 

\section{Implementation}

\subsection{Generating the free field on the torus}

To generate an instance of the free field on the torus $\mathbb{T}_n$, we adapt from the one-line pseudocode in Sheffield's survey paper \cite{SheffGFF}*{Section 4.4}. 
\begin{itemize}
\item For each $1\leq j \leq n$ and each $1\leq k \leq n$, generate two independent standard normal random variables $Z_{j,k}^{(1)}$ and $Z_{j,k}^{(2)}$. 
\item Form the matrix $\left(\tilde{X}(j,k)\right)_{\substack{1\leq j \leq m\\ 1\leq k \leq n}}$ whose entries read
$$\tilde{X}^c(j,k) = \sqrt{2\pi} \left(Z_{j,k}^{(1)} + \sqrt{-1}\cdot Z_{j,k}^{(2)}\right)\times \frac{\mathbbm{1}_{\{j+k\neq 2\}}}{2\sqrt{\sin^2\left(\frac{(j-1)\pi}{n} \right)+ \sin^2\left(\frac{(k-1)\pi}{n}\right)} }.$$
\item Taking $X={\rm Re}(\tilde{X})$ gives the (real-valued) discrete free field on $\mathbb{T}_n$.

\end{itemize}

\subsection{Solving the heat equation}

For each realization of the free field $X_n(\omega)$ on the torus $\mathbb{T}_n$ and a starting point $x\in \mathbb{T}_n$, we implement the Crank-Nicolson method \cite{CrankNicolson} to solve the heat equation 
\begin{equation}
\left\{\begin{array}{ll}
\partial_t u= \Delta_{M_n} (\omega) u & \text{on}~\mathbb{T}_n \times [0,T] \\ u(\cdot, 0)=\mathbbm{1}_x & \text{on}~\mathbb{T}_n 
\end{array}\right.,
\end{equation}
where $\Delta_{M_n}(\omega)$ is the Laplacian for LQG given in (\ref{eq:LQGDF}). The resulting solution give the heat kernel $(y,t) \mapsto p_t(x,y)$ associated with the Laplacian $\Delta_{M_n}$ (equivalently, the Markov process on $\mathbb{T}_n$ which is invariant with respect to the Liouville measure $M_n$).

In the results to follow we use $n=1024$.

\section{Results}

\subsection{On-diagonal heat kernel}

\begin{figure}[htp]
\centering
\includegraphics[scale=0.263]{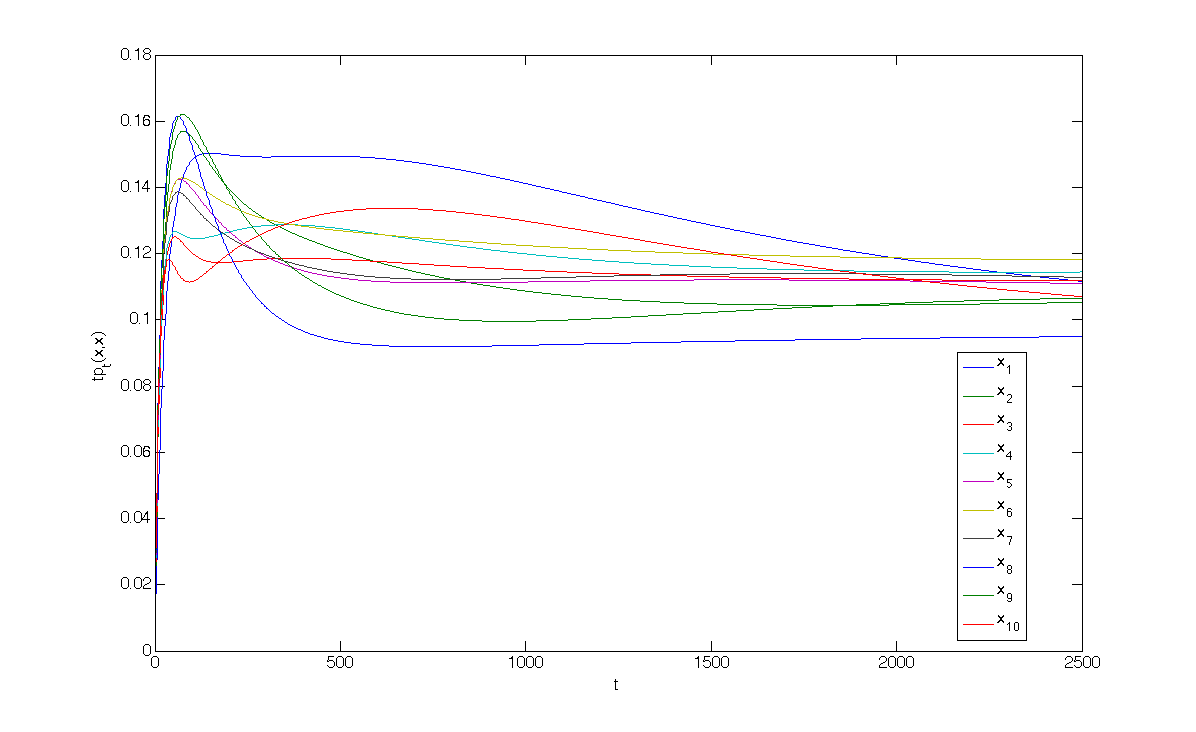}
\includegraphics[scale=0.263]{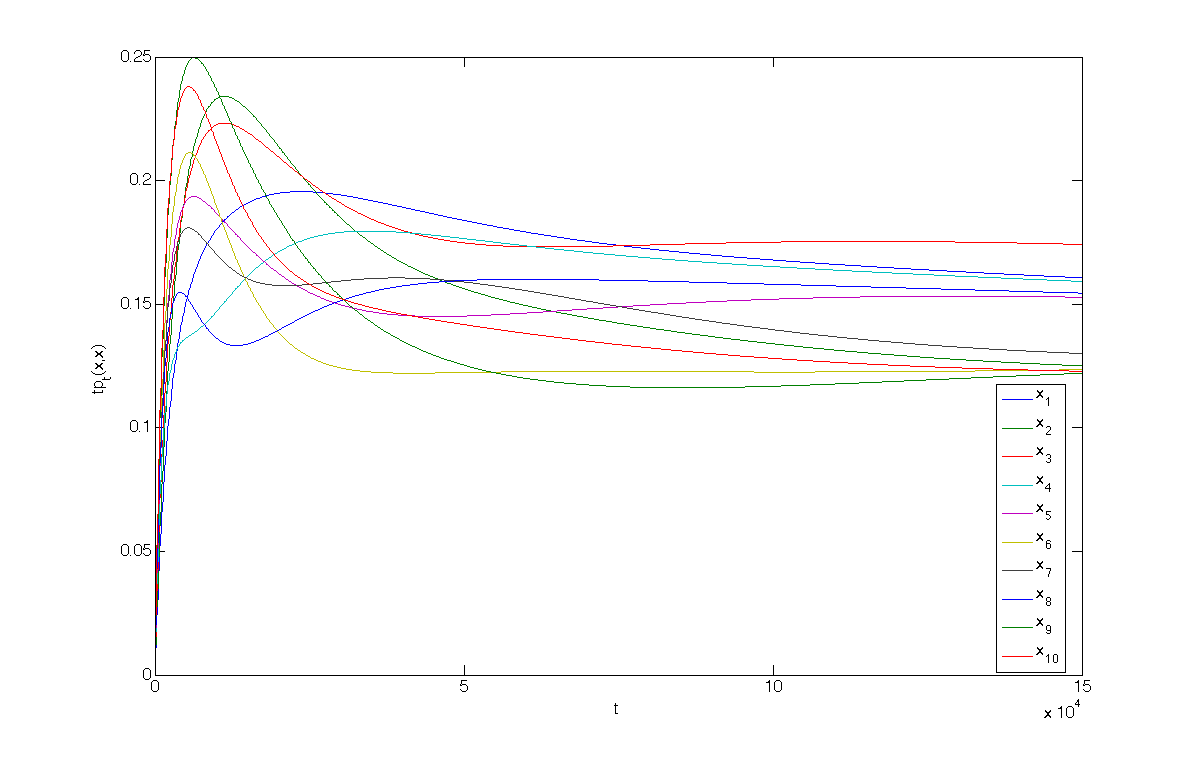}
\includegraphics[scale=0.263]{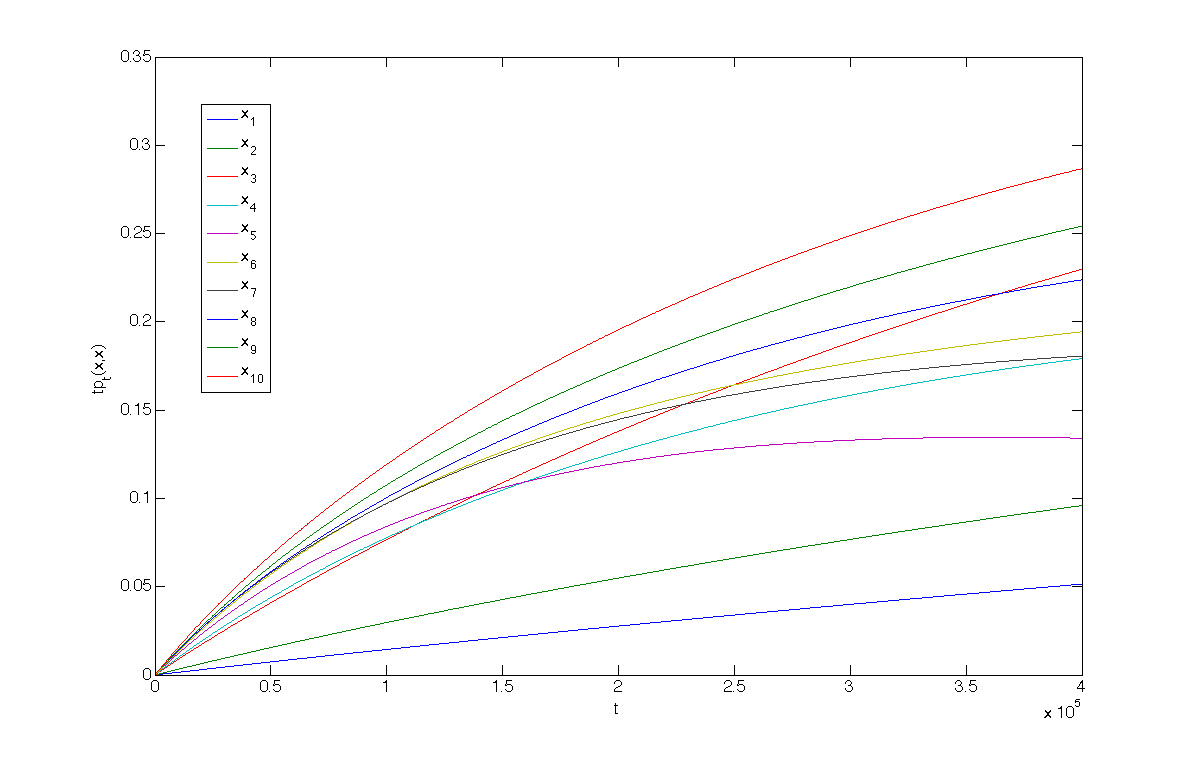}
\caption{On-diagonal heat kernel: $tp_t(x,x)$ vs. $t$ for the 10 highest points of some realizations of the free field. From top to bottom: $\gamma=0.4$, $\gamma=0.8$, and $\gamma=1.2$.}
\label{fig:OnDiagHK}
\end{figure}

In Figure \ref{fig:OnDiagHK} we plot $t \cdot p_t(x,x)$ versus time $t$ for various starting points $x\in \mathbb{T}$. Our numerical computations support the on-diagonal asymptotics $p_t(x,x) \asymp t^{-1}$ for moderately short times. Note that due to the discretization of the lattice, $p_t(x,x)$ remains bounded as $t\downarrow 0$, so we do not interpret what happens in the very short time regime. 

\subsection{Off-diagonal heat kernel: heat balls}

\begin{figure}[htp]
\centering
\includegraphics[scale=0.25]{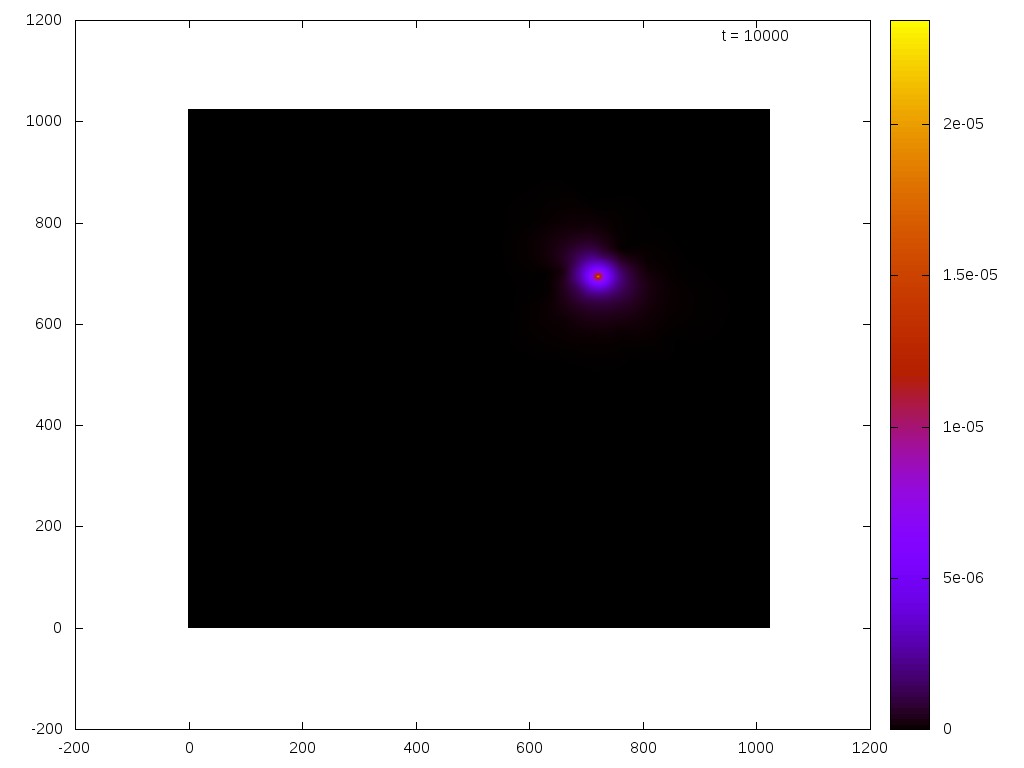}
\includegraphics[scale=0.25]{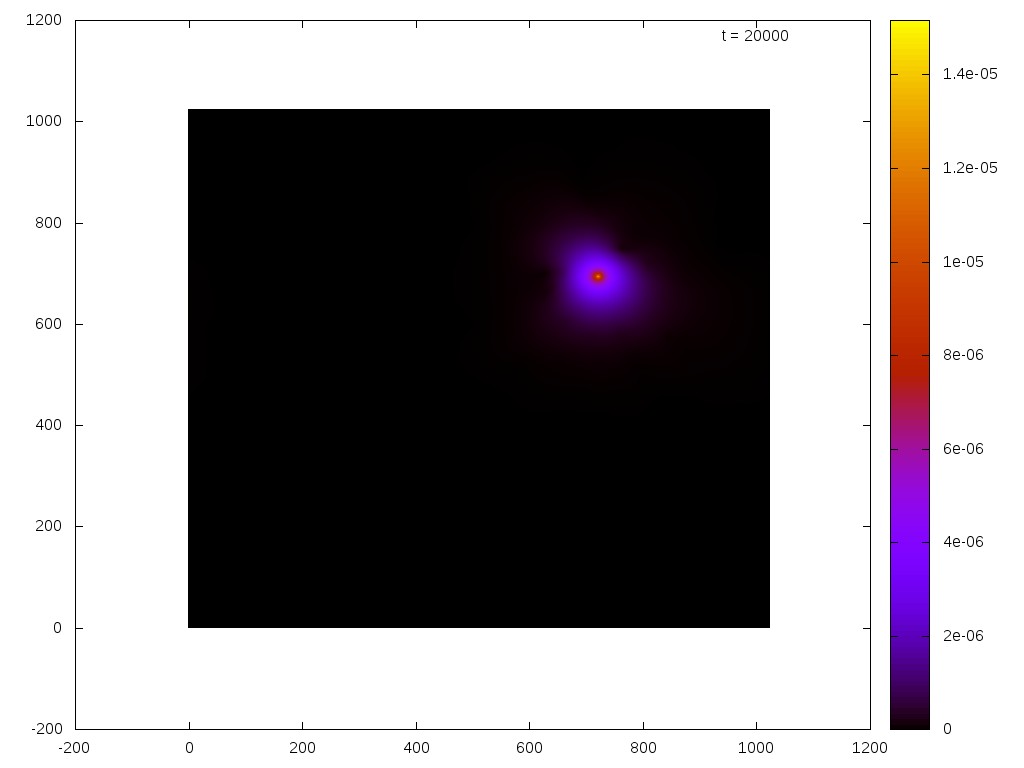}
\includegraphics[scale=0.25]{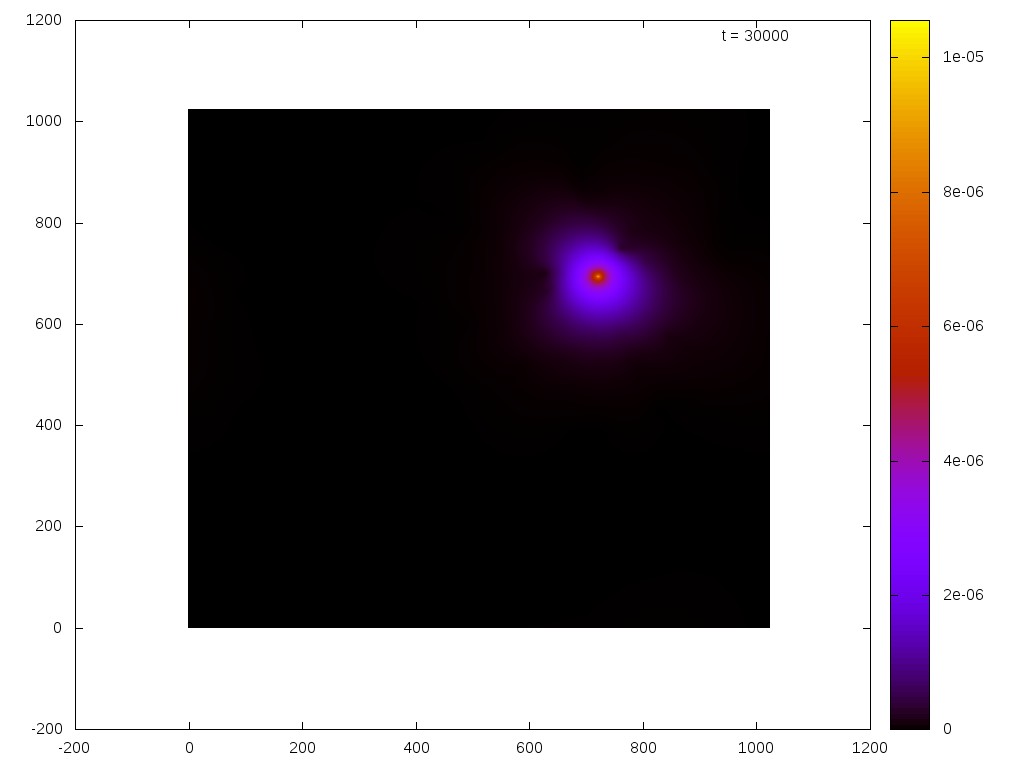}
\includegraphics[scale=0.25]{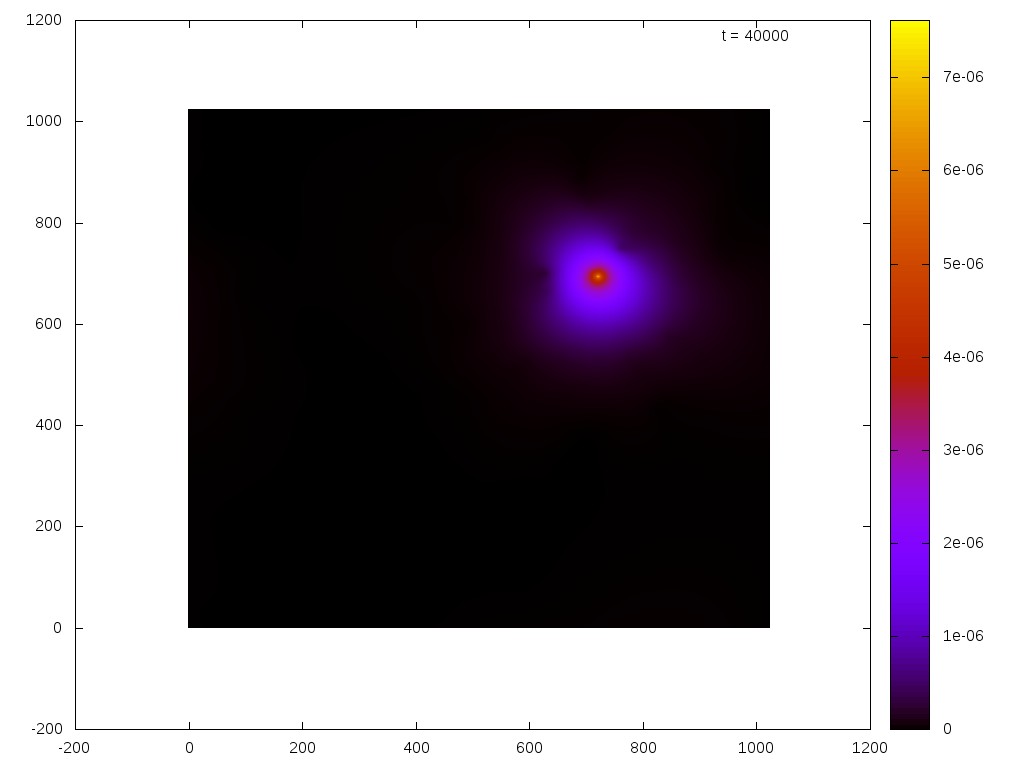}
\includegraphics[scale=0.25]{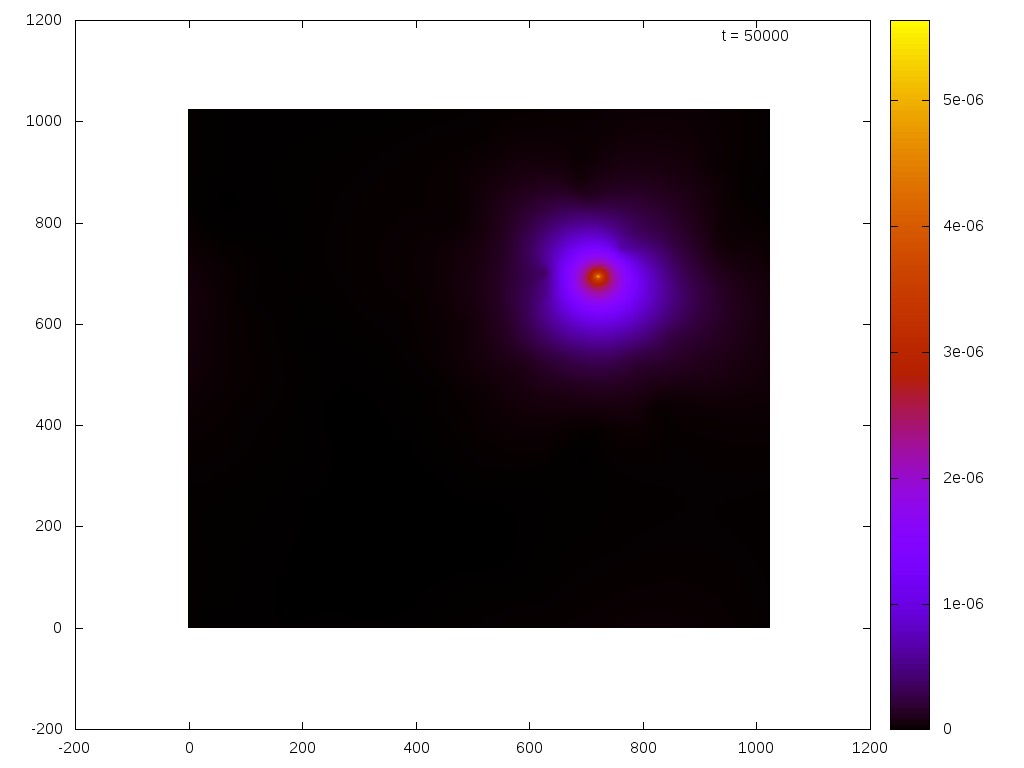}
\includegraphics[scale=0.25]{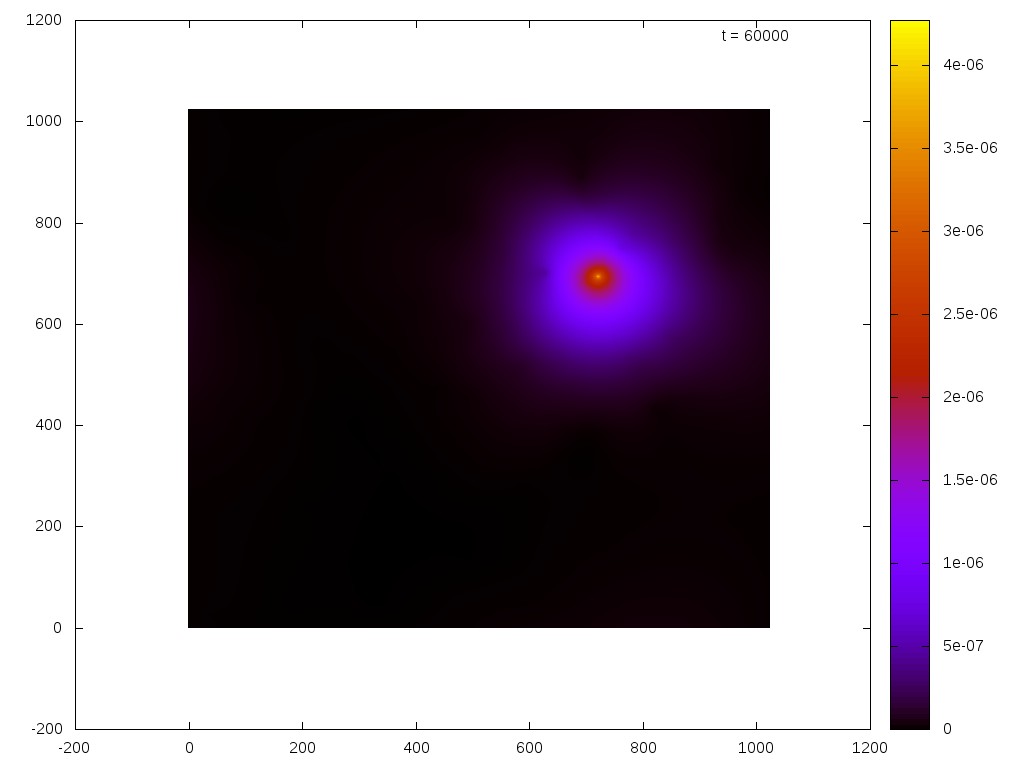}
\includegraphics[scale=0.25]{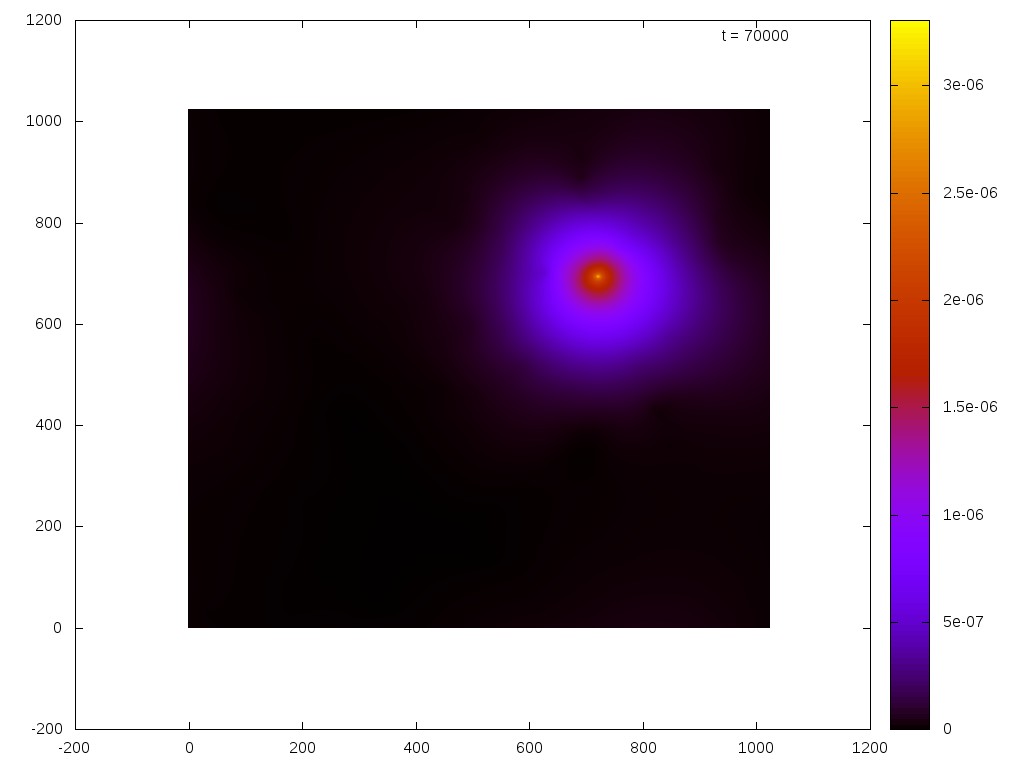}
\includegraphics[scale=0.25]{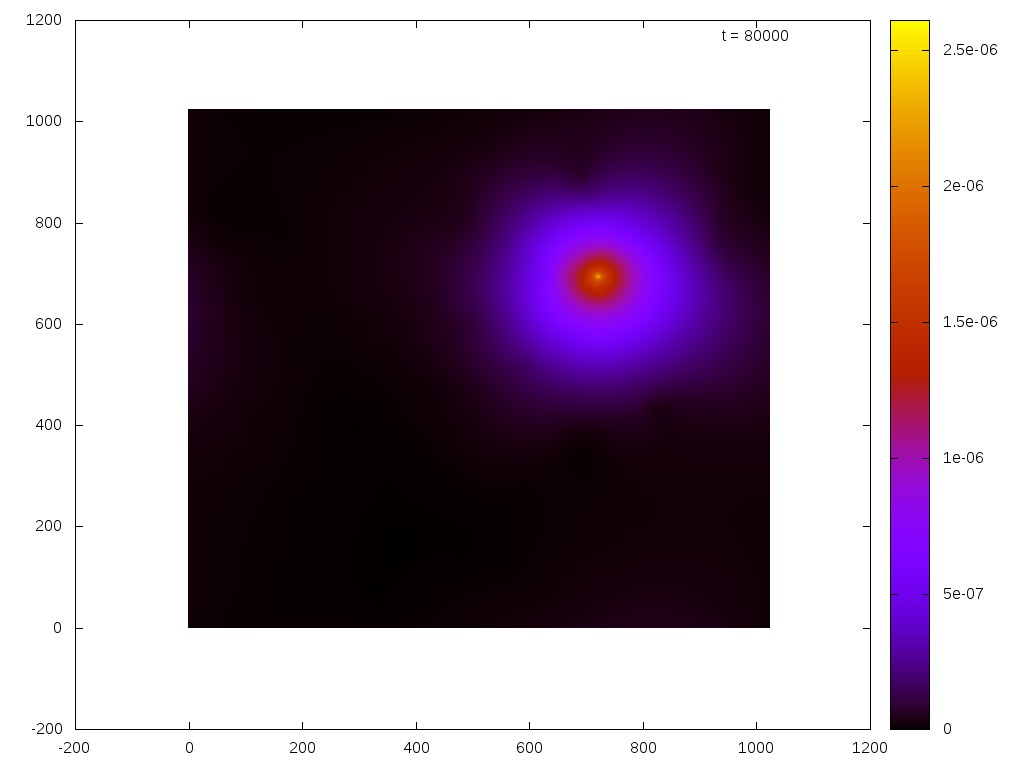}
\caption{(Color online.) Snapshots of the heat kernel started from the highest point of a particular free field realization, $\gamma=0.8$. Each snapshot is produced at time steps of $10,000$ iterations of the Crank-Nicolson method.}
\label{fig:heatballs}
\end{figure}

\begin{figure}[htp]
\centering
\includegraphics[scale=0.25]{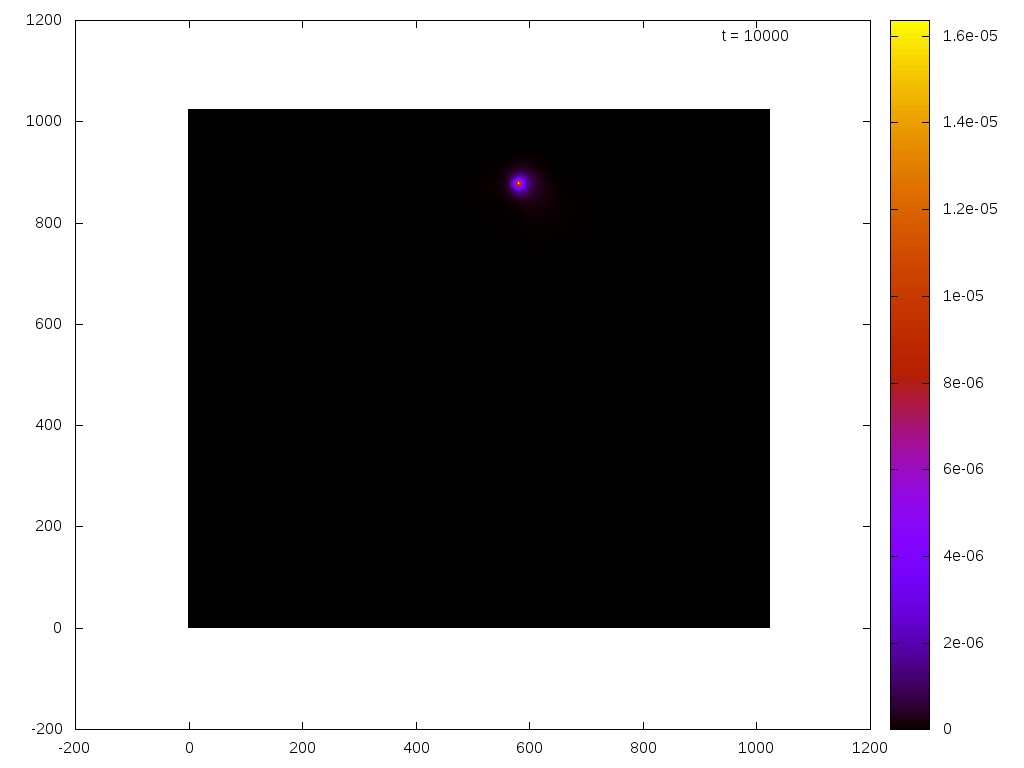}
\includegraphics[scale=0.25]{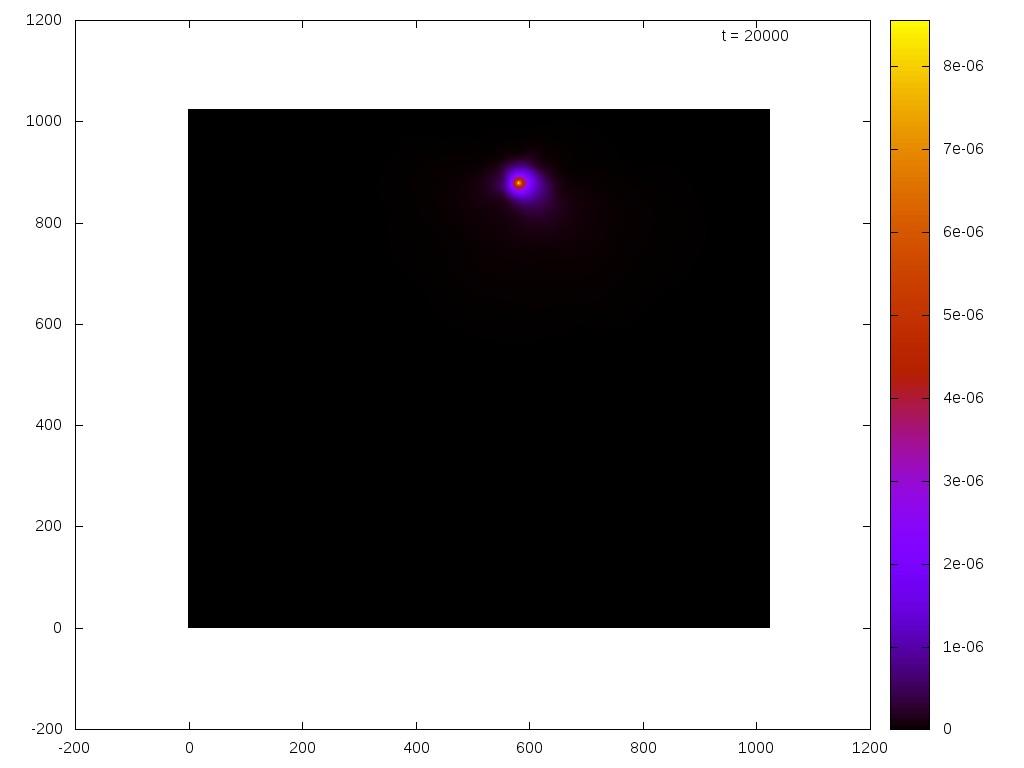}
\includegraphics[scale=0.25]{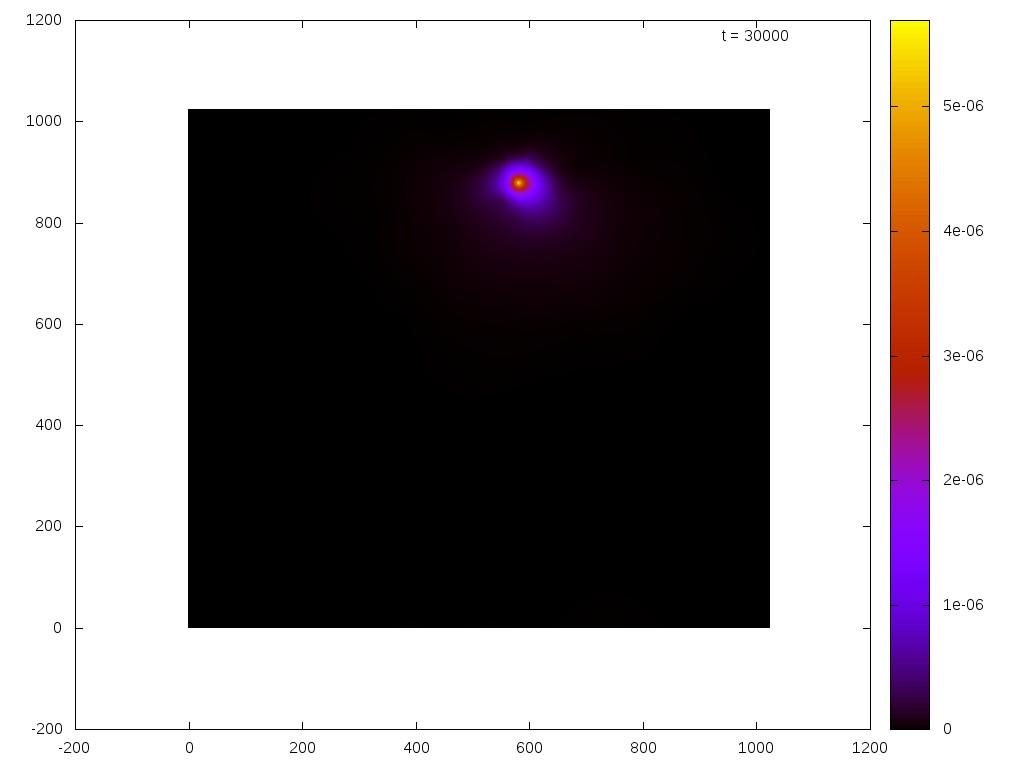}
\includegraphics[scale=0.25]{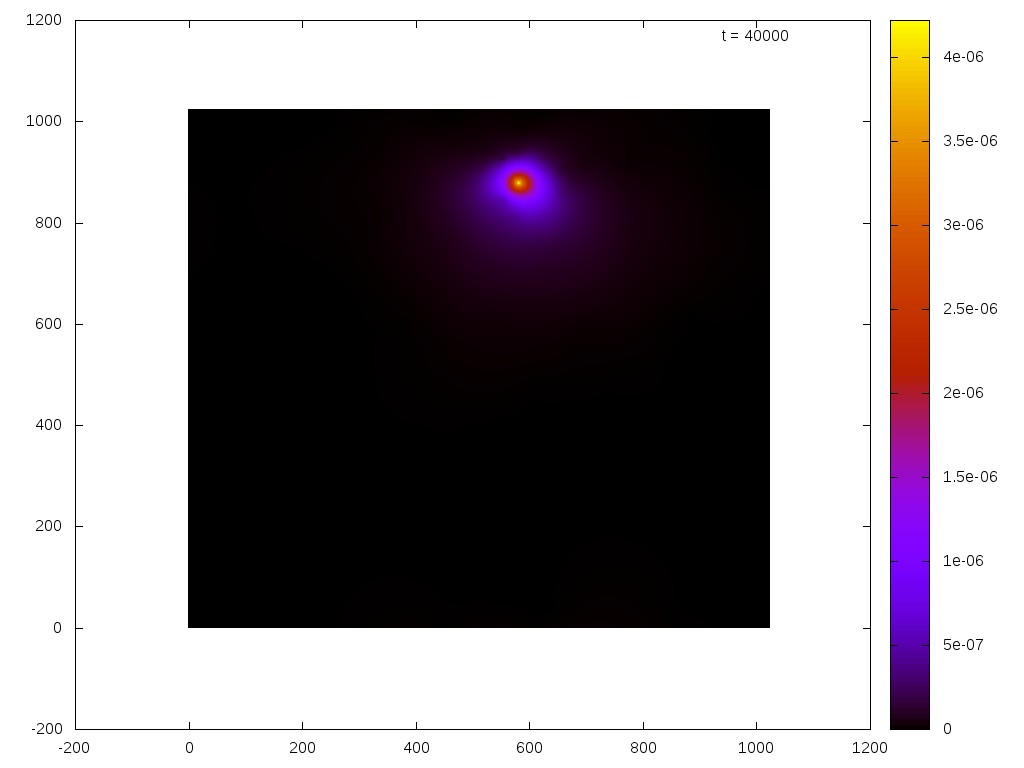}
\includegraphics[scale=0.25]{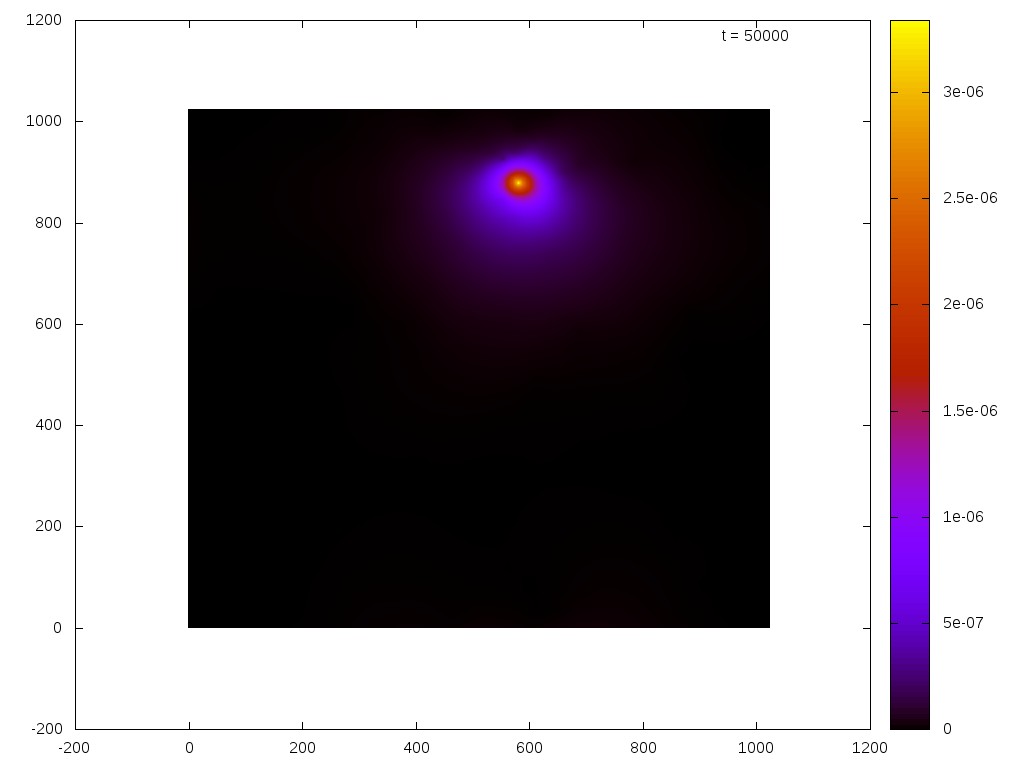}
\includegraphics[scale=0.25]{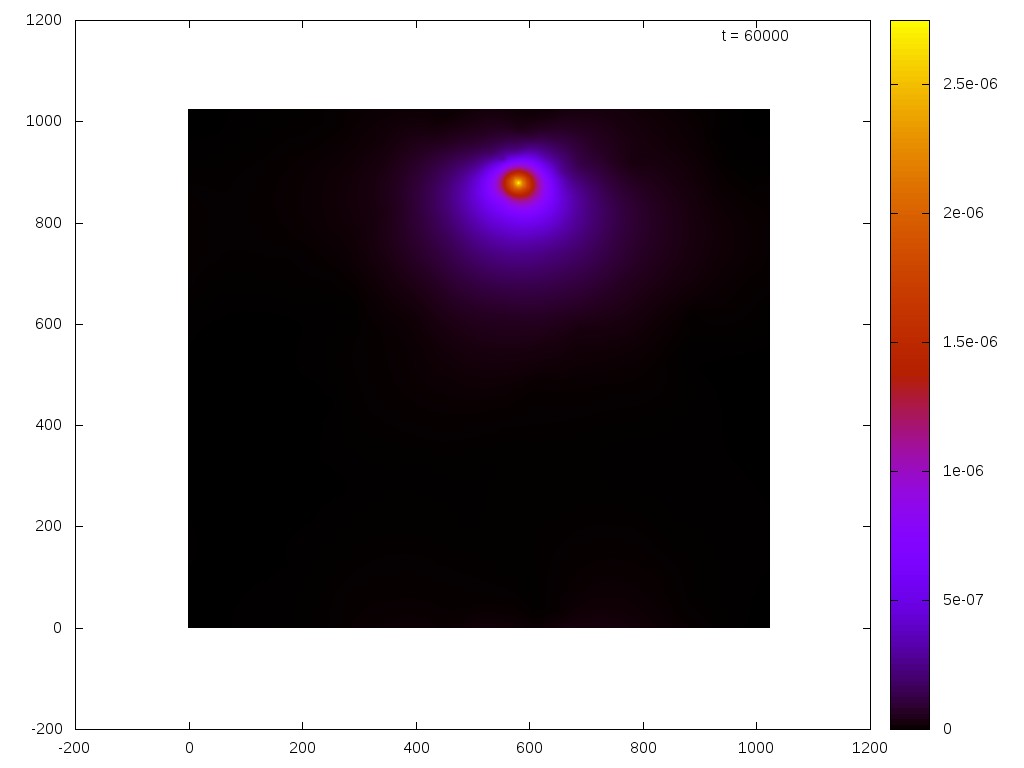}
\includegraphics[scale=0.25]{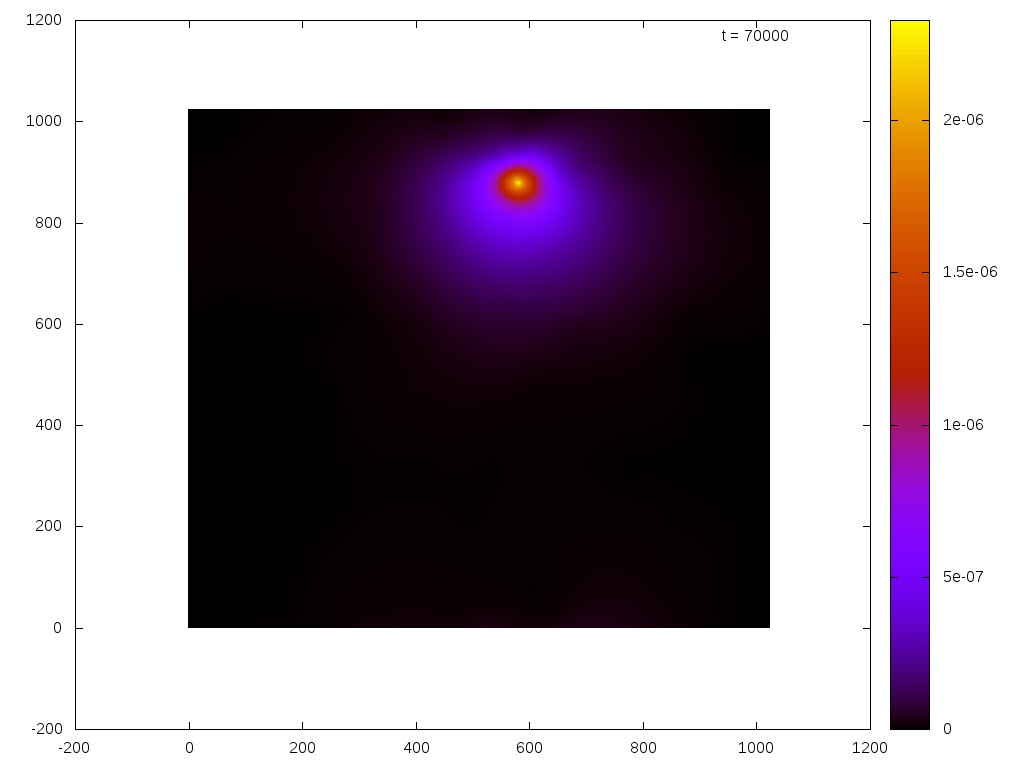}
\includegraphics[scale=0.25]{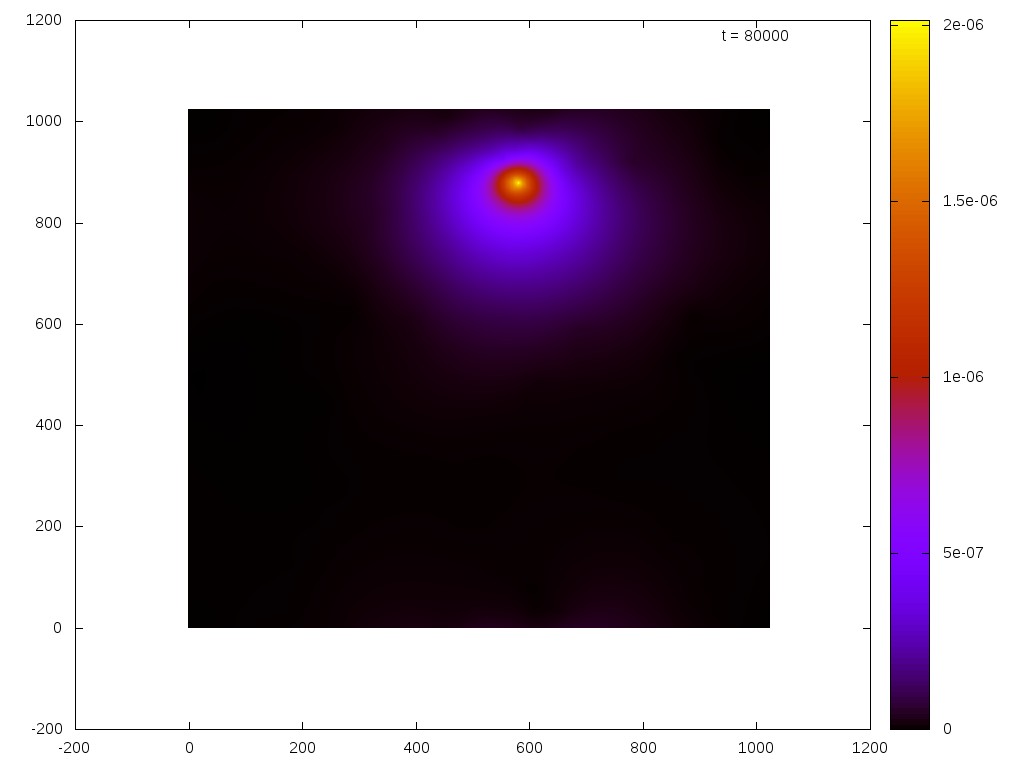}
\caption{(Color online.) Snapshots of the heat kernel started from the 3rd highest point of a different free field realization, $\gamma=0.8$. Each snapshot is produced at time steps of $10,000$ iterations of the Crank-Nicolson method.}
\label{fig:heatballs2}
\end{figure}

Now we turn to the off-diagonal heat kernel $p_t(x,y)$. Figures \ref{fig:heatballs} and \ref{fig:heatballs2} show the heat kernels $y\mapsto p_t(x,y)$ for several ``snapshots'' taken at regular  intervals of $t$. Colors reveal the profiles of the ``heat balls'' as diffusion proceeds in time. Qualitatively, when $x$ is a high point of the free field $X_n$, the heat ball grows more slowly in time, and its shape stays isotropic for most of the time. When $x$ is not a high point, the speed of diffusion is faster, and the shape of the heat ball is more prone to distortion from nearby obstacle points, \emph{i.e.,} points $z$ where $X_n(z)  > X_n(x)$. These observations appear to hold without regard to the particular realizations of the free field.

\subsection{Off-diagonal heat kernel: short-time behavior}

 In order to extract the dependence of the heat kernel on the space-time parameter $|x-y|^\alpha/t$, we plot the quantity $p_t(x,y)/p_t(x,x)$ as a function of $|x-y|^\alpha/t$, for several values of $t$. Here $\alpha$ is a positive number to be determined. Let us recall that in the Euclidean setting ($\gamma=0$), the heat kernel takes the form
\begin{equation}
\label{eq:EucHK}
p_t(x,y) = \frac{1}{4\pi t}\exp\left(-\frac{|x-y|^2}{4t}\right)\qquad \forall x,y \in \mathbb{R}^2,
\end{equation}
which means that $\alpha=2$ and $p_t(x,y)/p_t(x,x) = \exp(-|x-y|^2/(4t))$. If, for a fixed $x$, one makes the graphs of $p_t(x,y)/p_t(x,x)$ versus $|x-y|^2/t$ for several values of $t$, then according to (\ref{eq:EucHK}), the curves for different times $t$ ought to collapse onto one single curve. 

\begin{figure}[htp]
\centering
\includegraphics[scale=0.2]{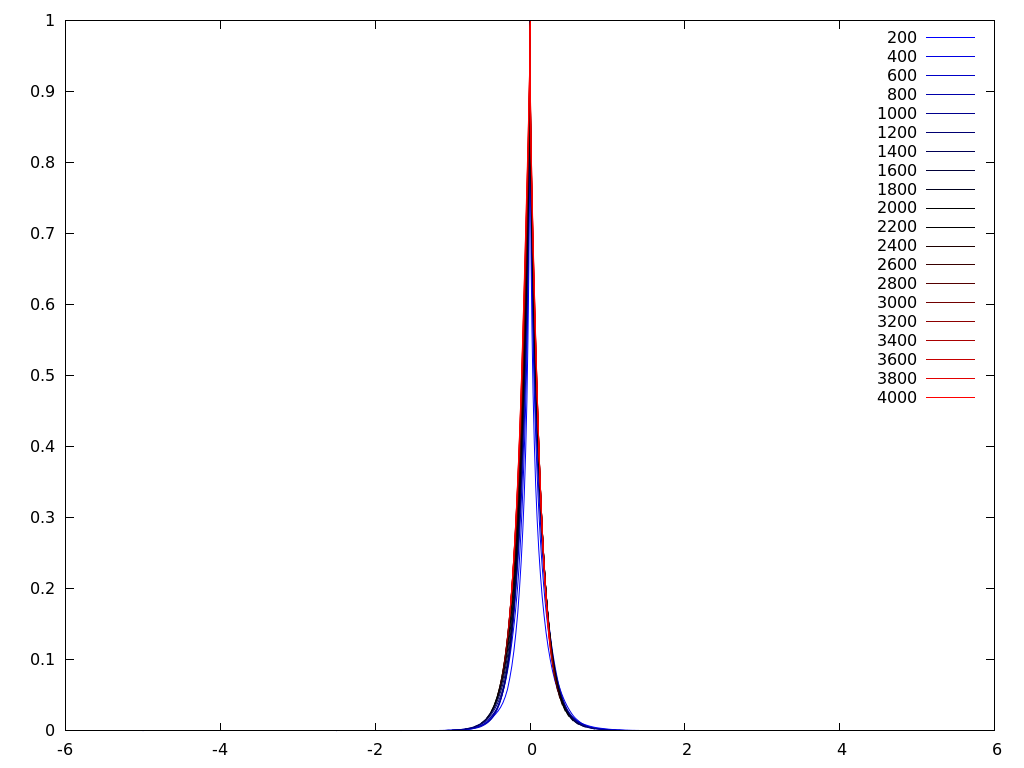}
\includegraphics[scale=0.2]{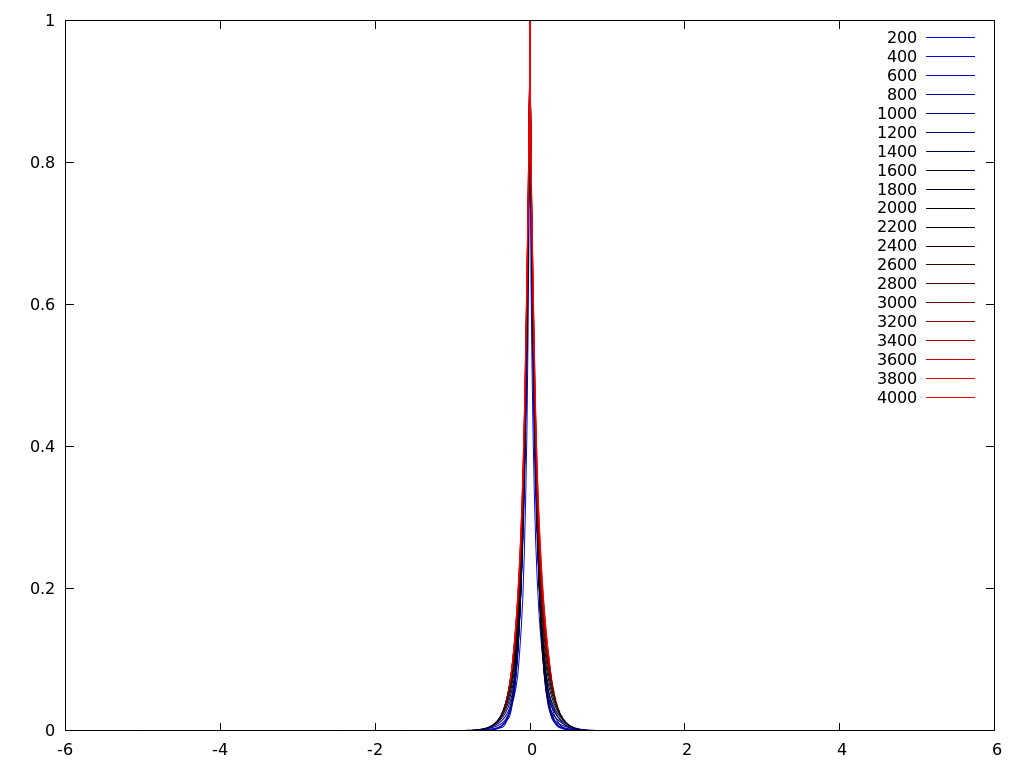}
\includegraphics[scale=0.2]{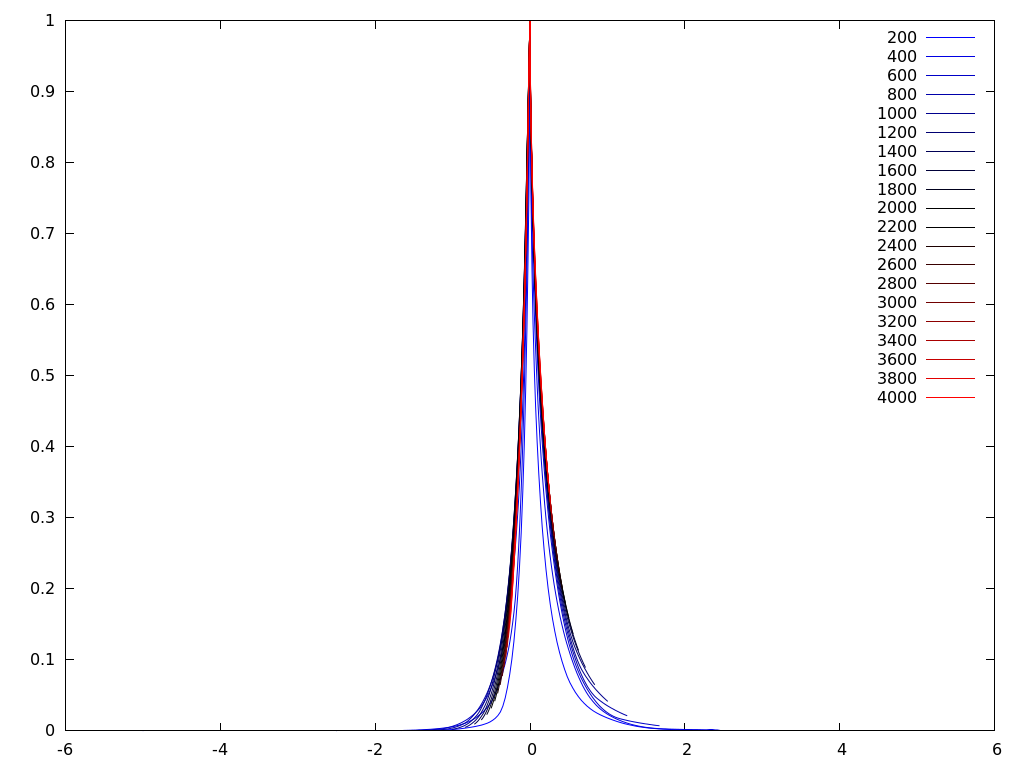}
\includegraphics[scale=0.2]{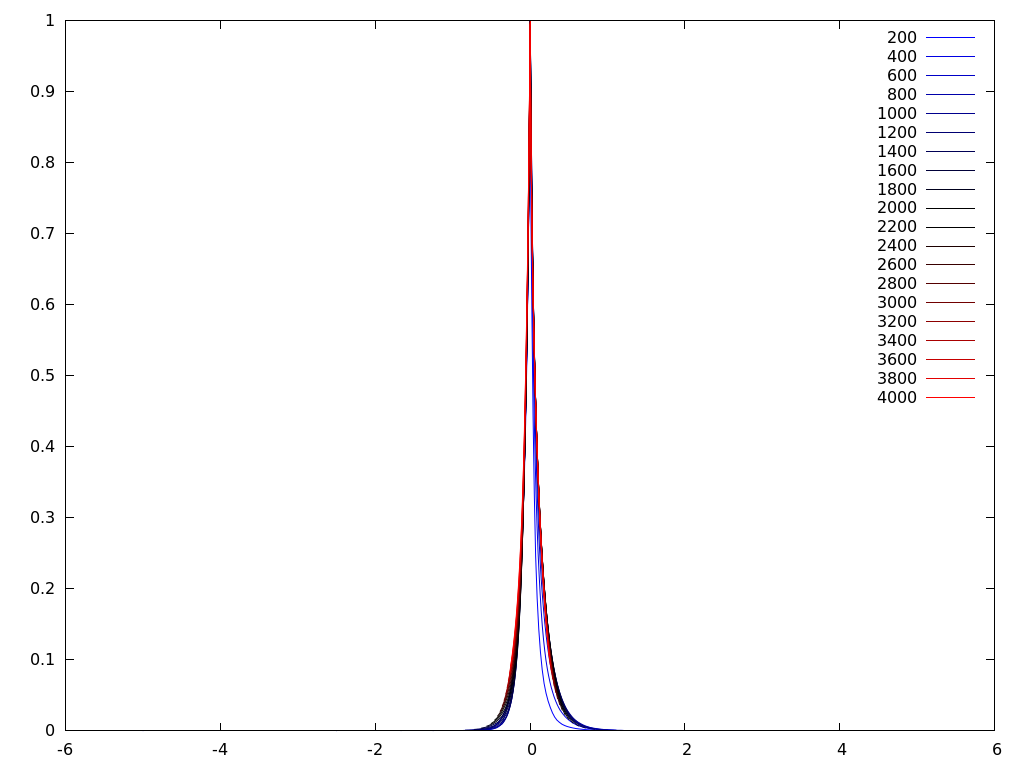}
\includegraphics[scale=0.2]{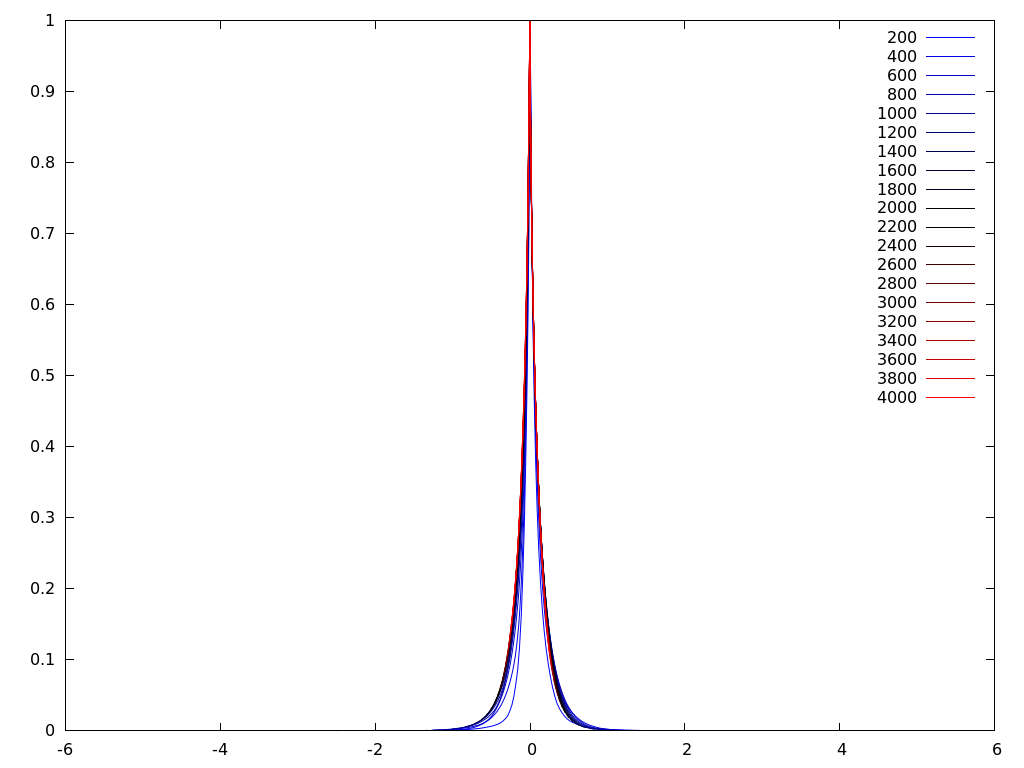}
\includegraphics[scale=0.2]{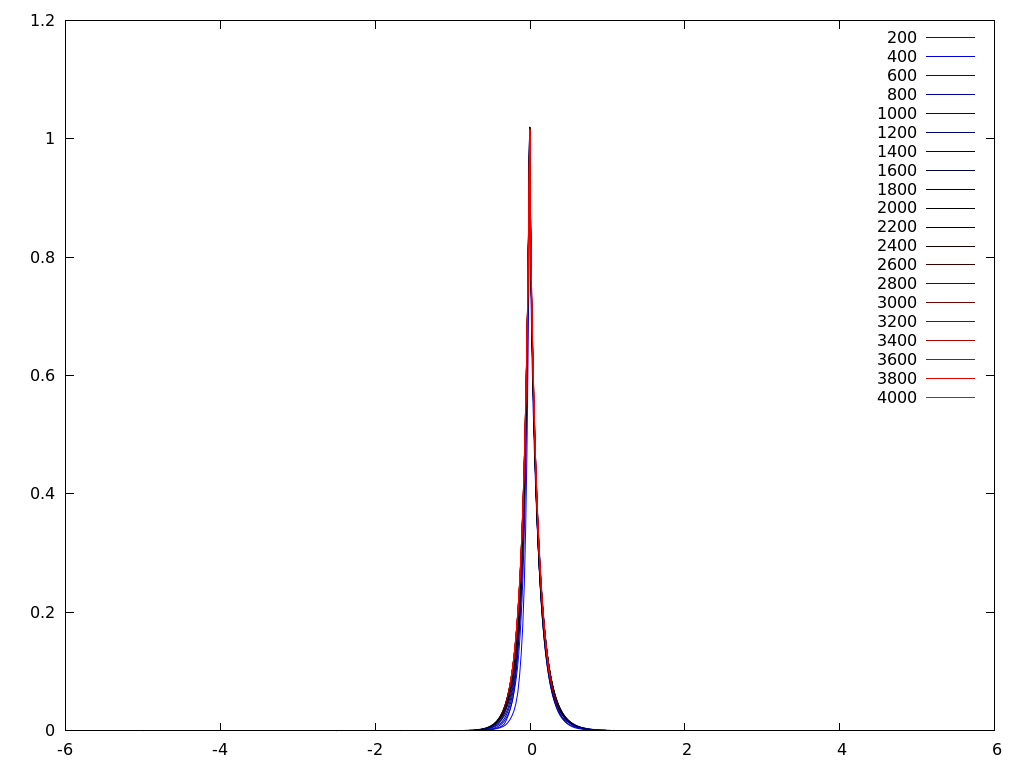}
\caption{The off-diagonal heat kernel $p_t(x,y)/p_t(x,x)$ plotted versus $|x-y|^\alpha/t$, where $x$ is each of the 6 highest points of a particular free field realization ($\gamma=0.4$, $\alpha=1.5$). The various times $t$, measured in the number of iterations of the Crank-Nicolson method, are indicated in the legend.}
\label{fig:04}
\end{figure}

\begin{figure}[htp]
\centering
\includegraphics[scale=0.2]{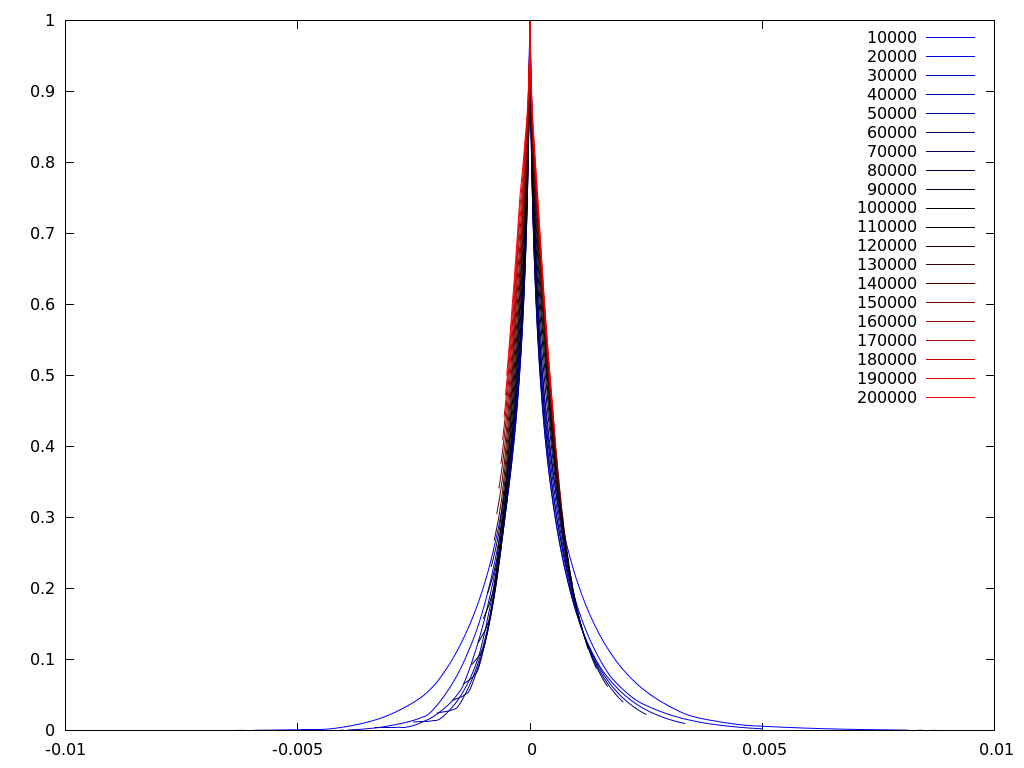}
\includegraphics[scale=0.2]{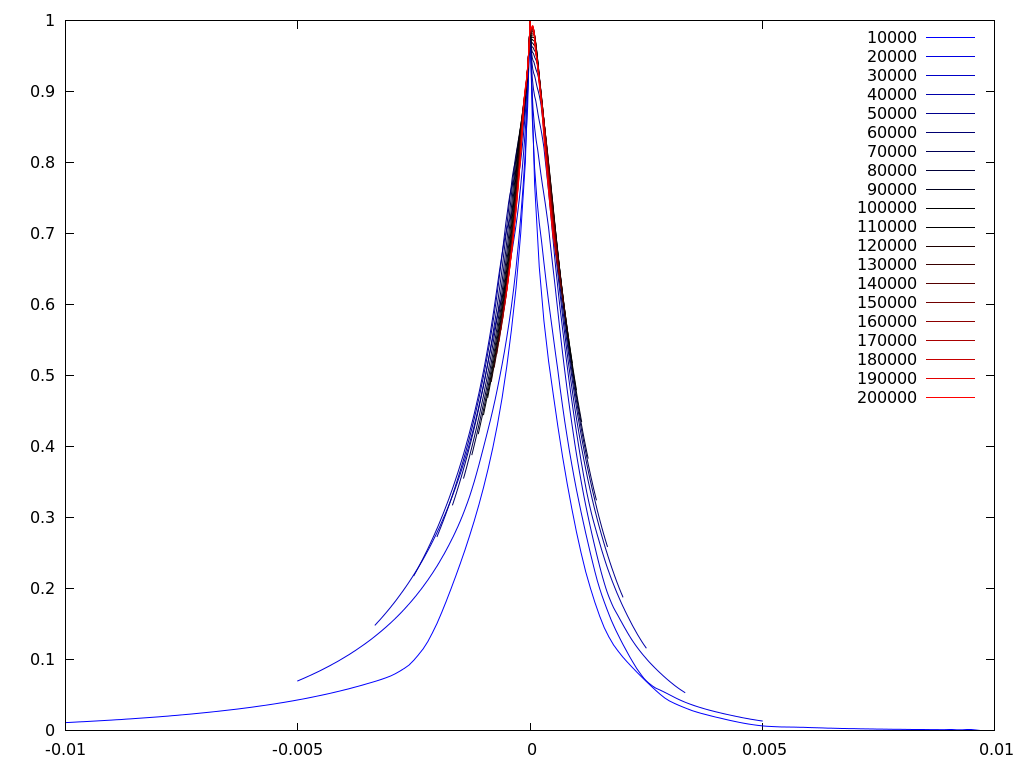}
\includegraphics[scale=0.2]{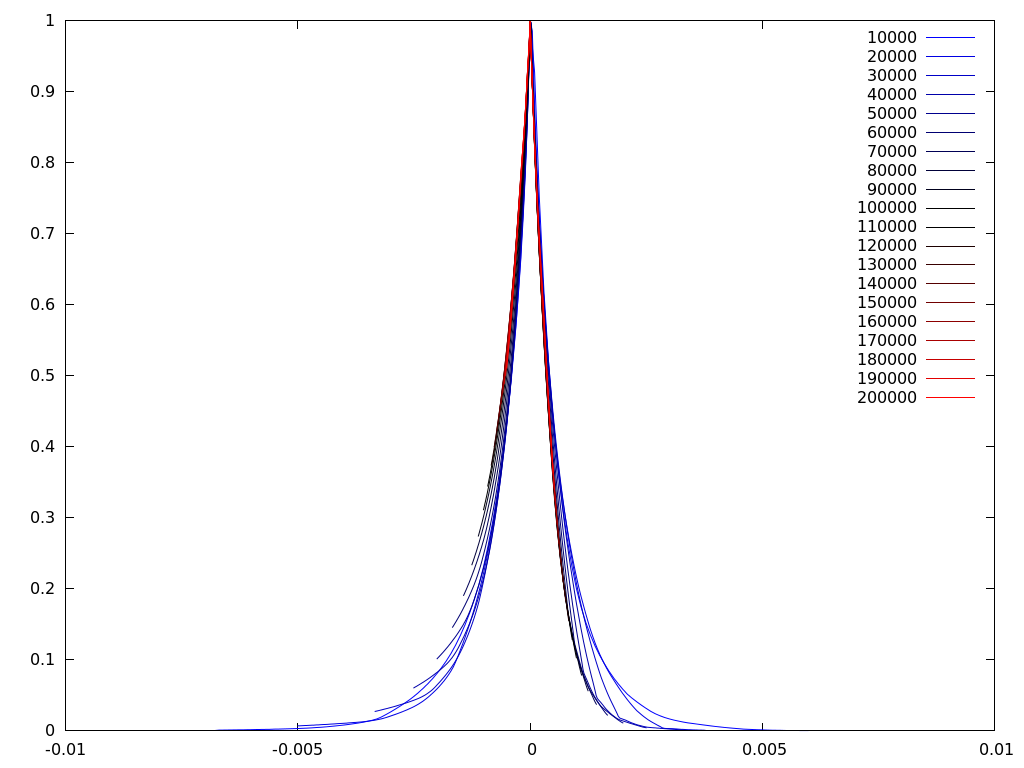}
\includegraphics[scale=0.2]{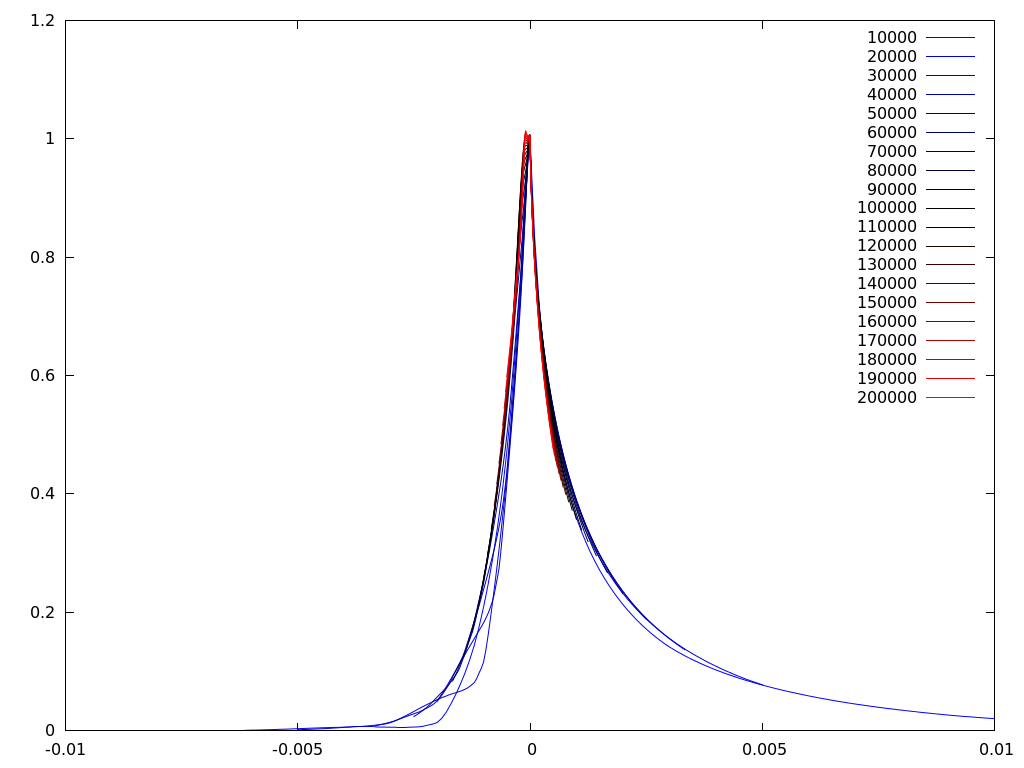}
\includegraphics[scale=0.2]{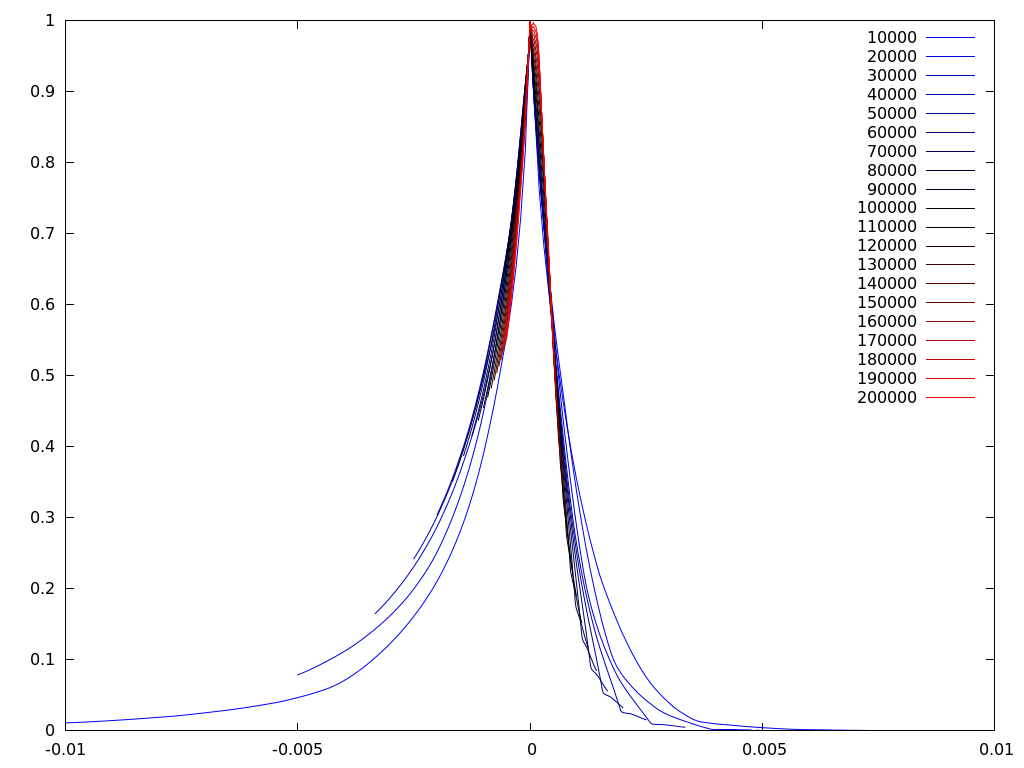}
\includegraphics[scale=0.2]{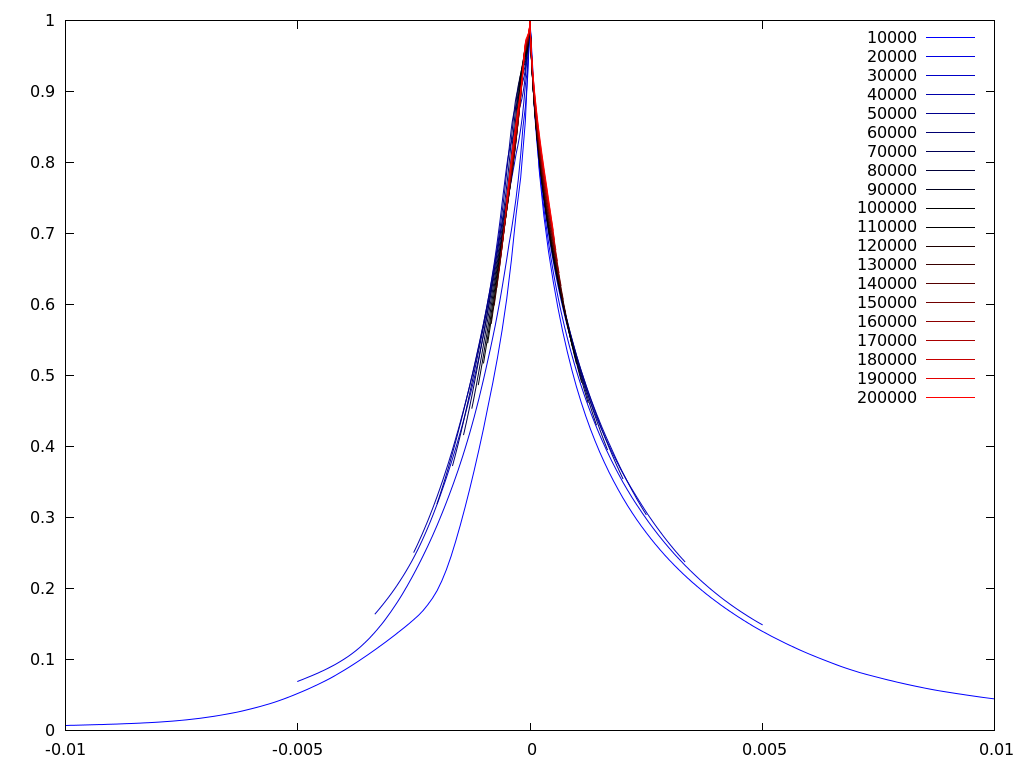}
\caption{The off-diagonal heat kernel $p_t(x,y)/p_t(x,x)$ plotted versus $|x-y|^\alpha/t$, where $x$ is each of the 6 highest points of a particular free field realization ($\gamma=0.8$, $\alpha=1$). The various times $t$, measured in the number of iterations of the Crank-Nicolson method, are indicated in the legend.}
\label{fig:08}
\end{figure}

\begin{figure}[htp]
\centering
\includegraphics[scale=0.2]{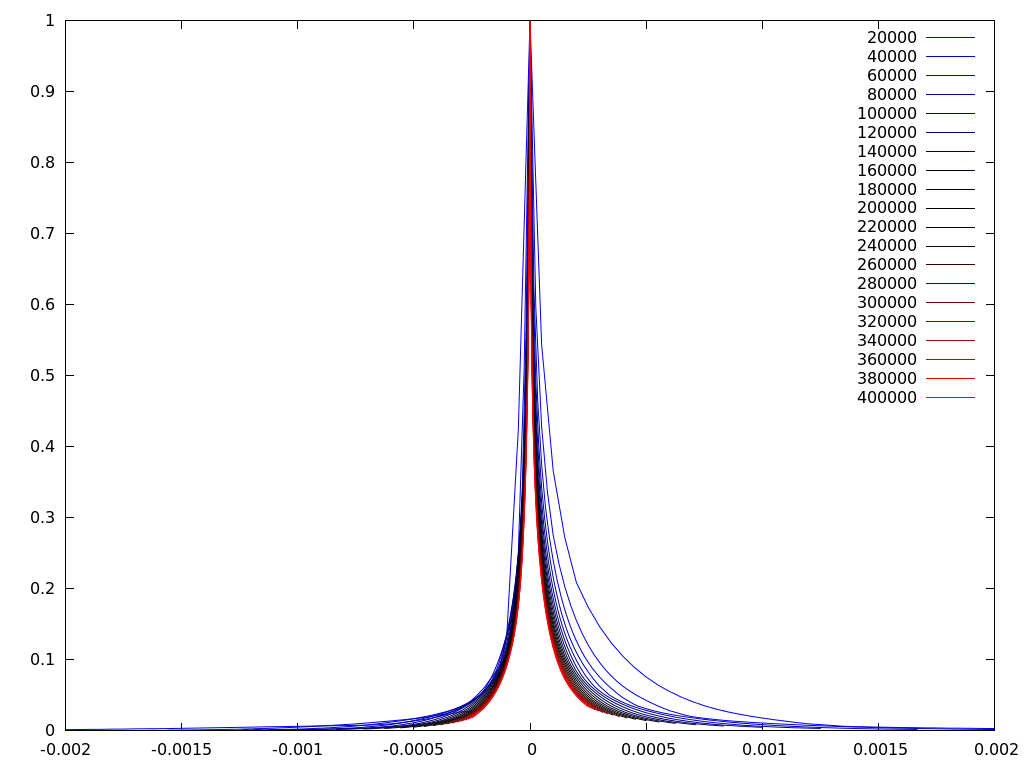}
\includegraphics[scale=0.2]{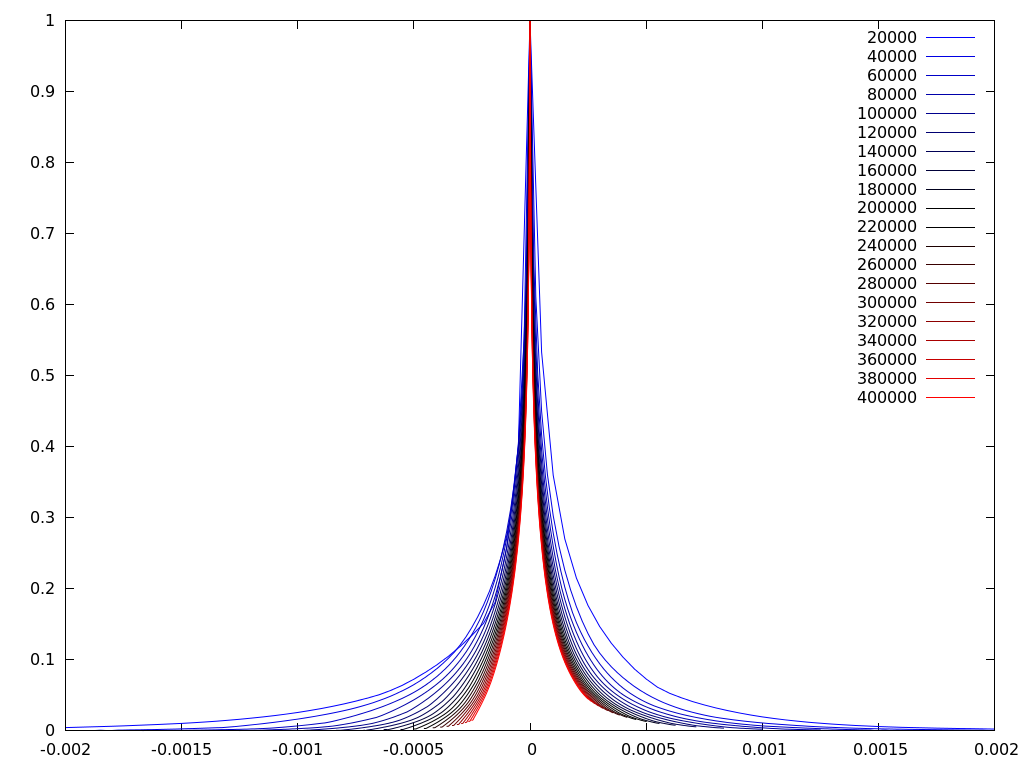}
\includegraphics[scale=0.2]{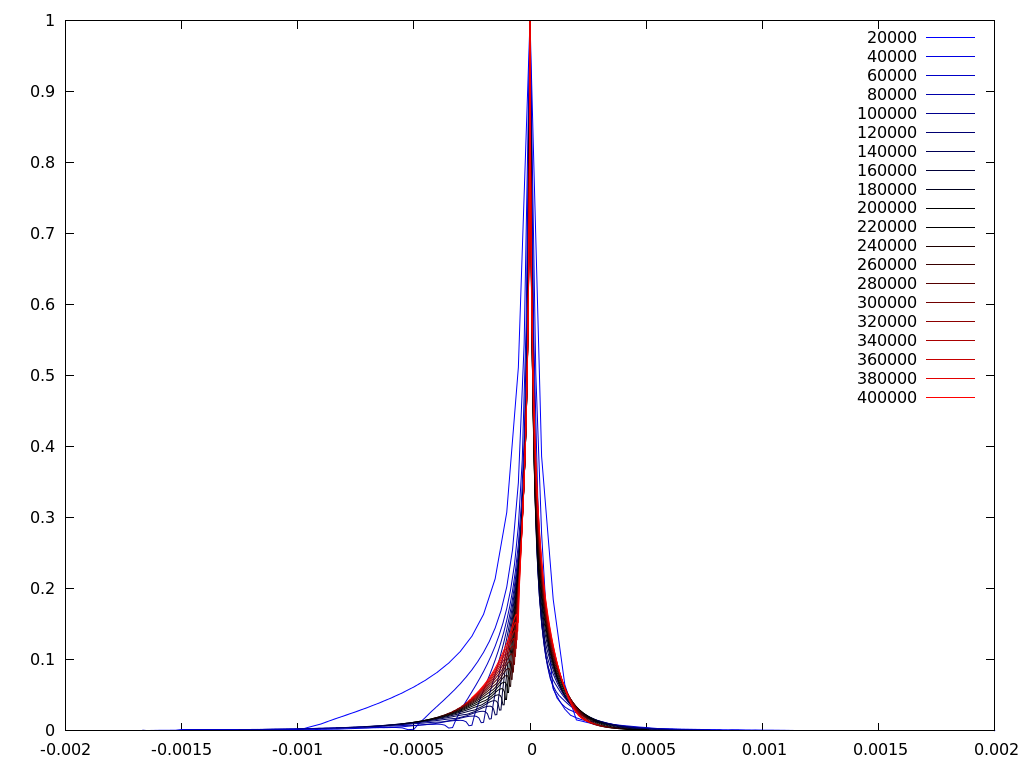}
\includegraphics[scale=0.2]{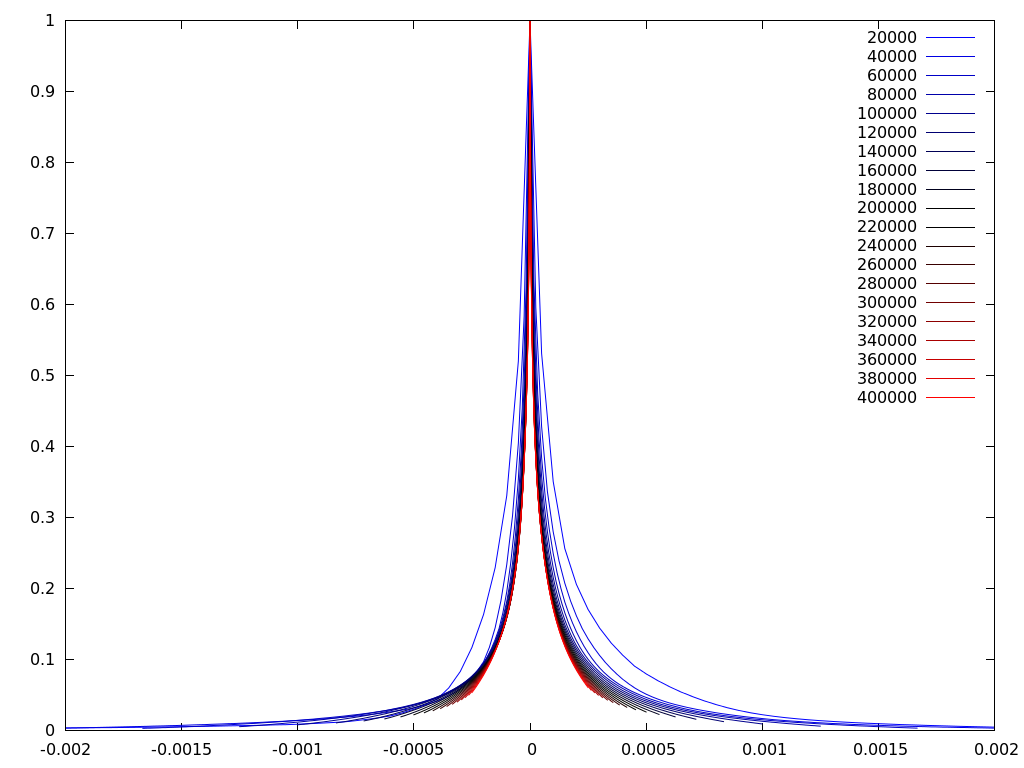}
\includegraphics[scale=0.2]{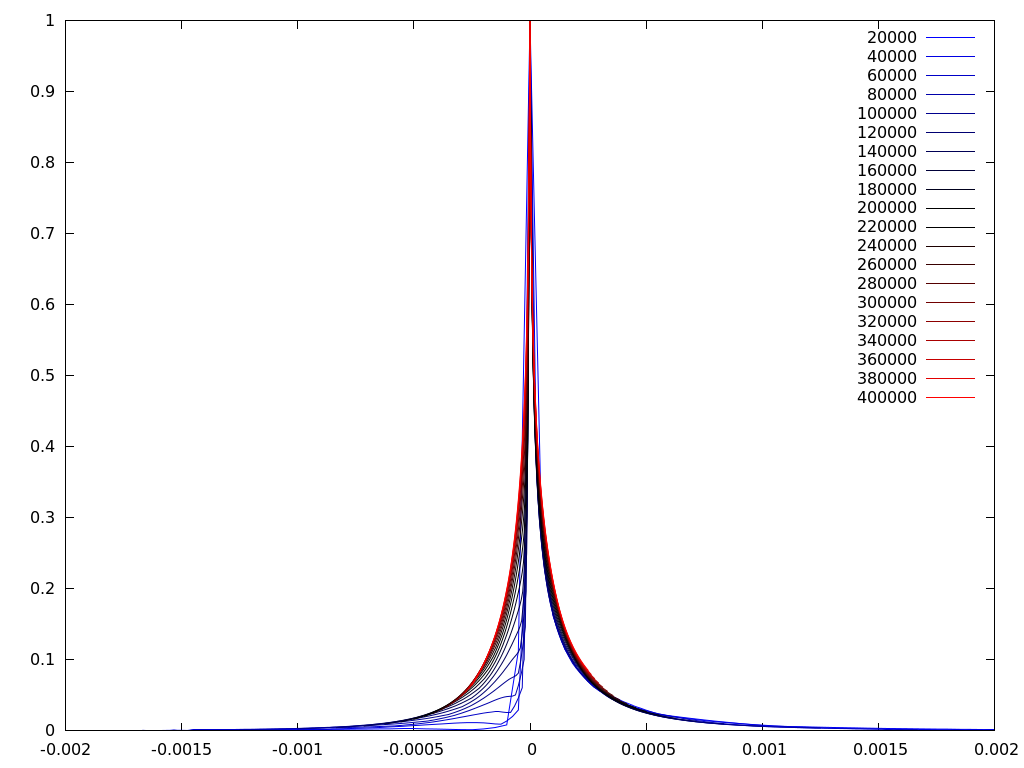}
\includegraphics[scale=0.2]{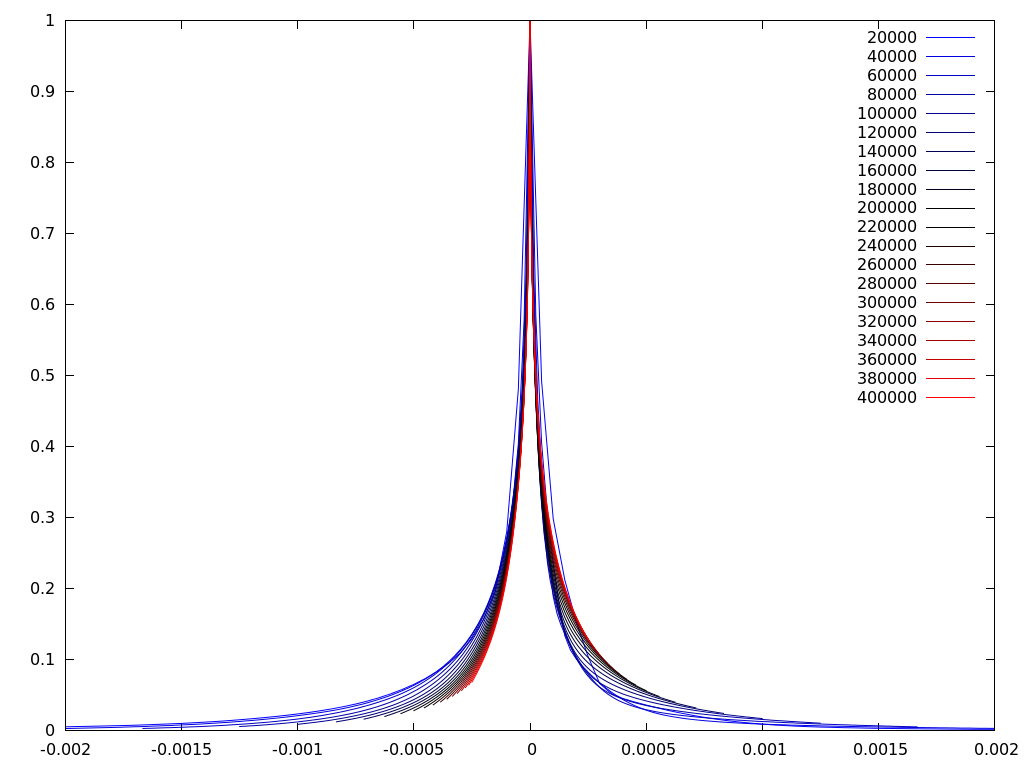}
\caption{The off-diagonal heat kernel $p_t(x,y)/p_t(x,x)$ plotted versus $|x-y|^\alpha/t$, where $x$ is each of the 6 highest points of a particular free field realization ($\gamma=1.2$, $\alpha=1$). The various times $t$, measured in the number of iterations of the Crank-Nicolson method, are indicated in the legend.}
\label{fig:12}
\end{figure}

\begin{figure}[htp]
\centering
\includegraphics[scale=0.2]{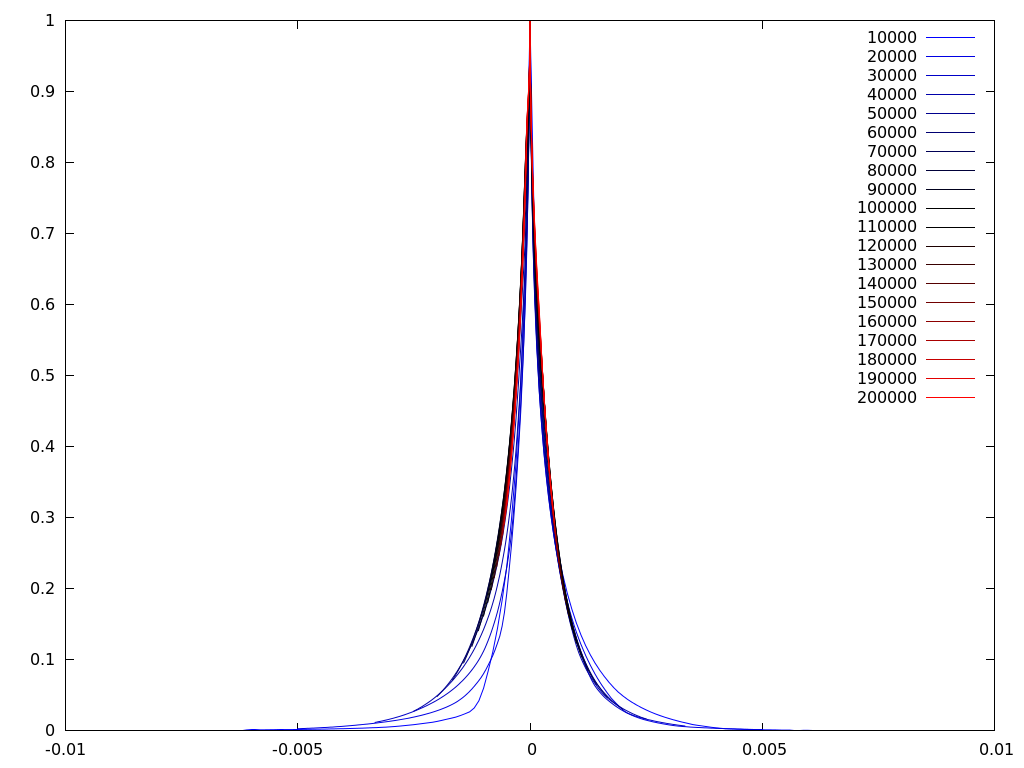}
\includegraphics[scale=0.2]{alpha-seed2-high01}
\includegraphics[scale=0.2]{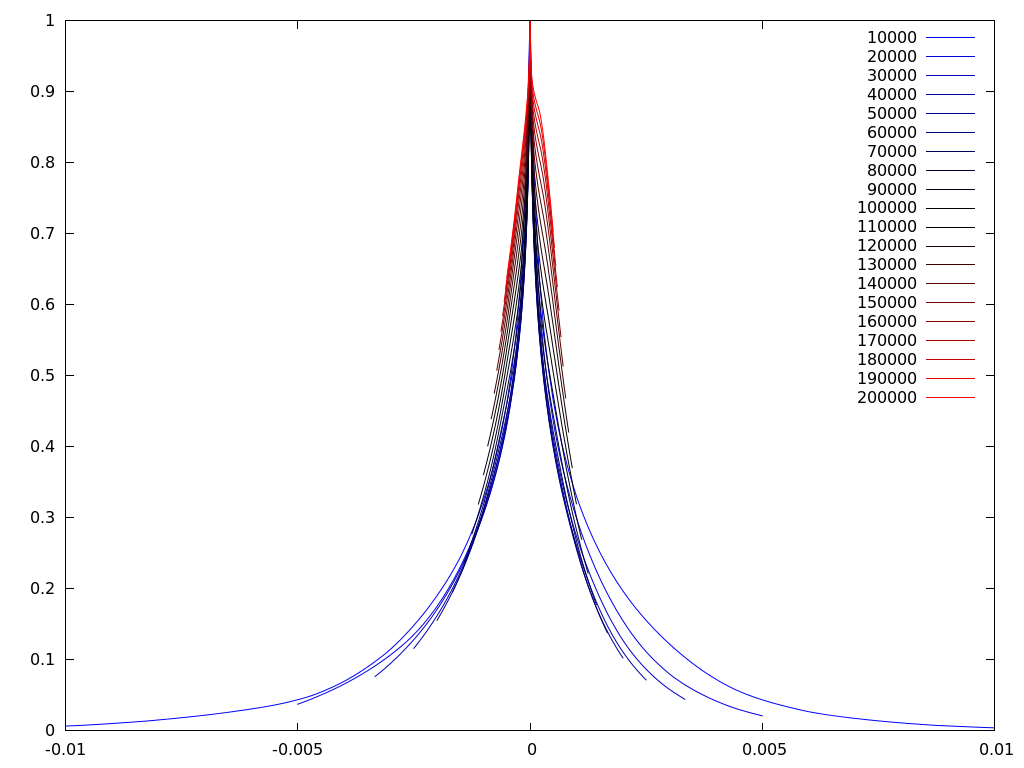}
\includegraphics[scale=0.2]{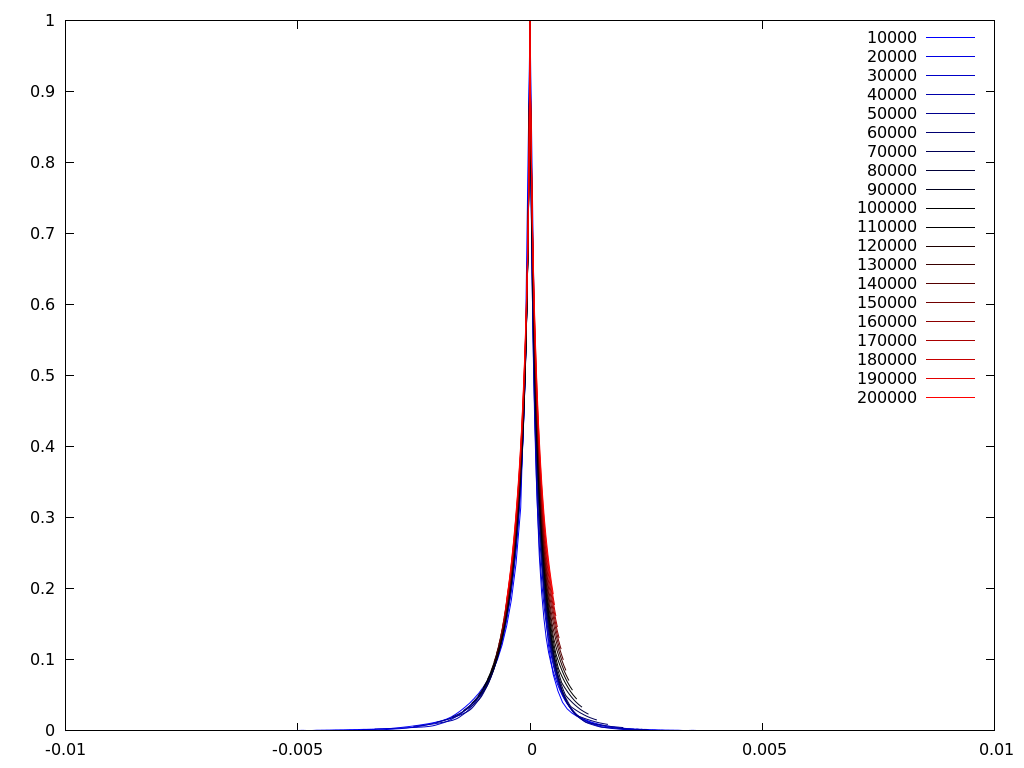}
\includegraphics[scale=0.2]{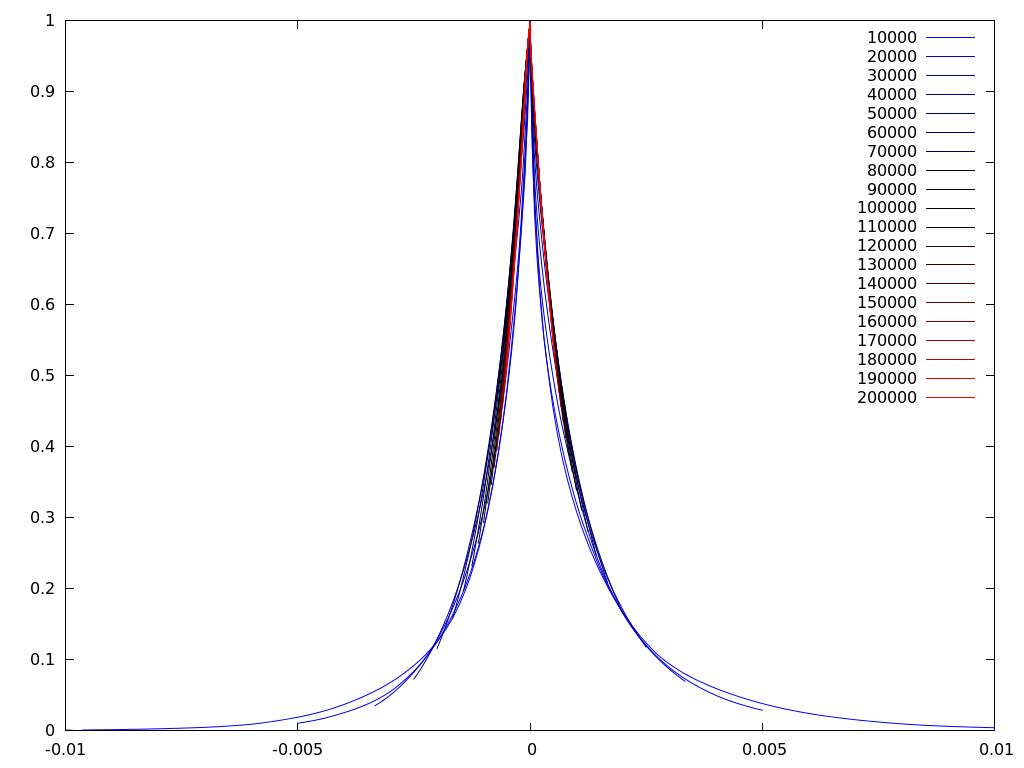}
\includegraphics[scale=0.2]{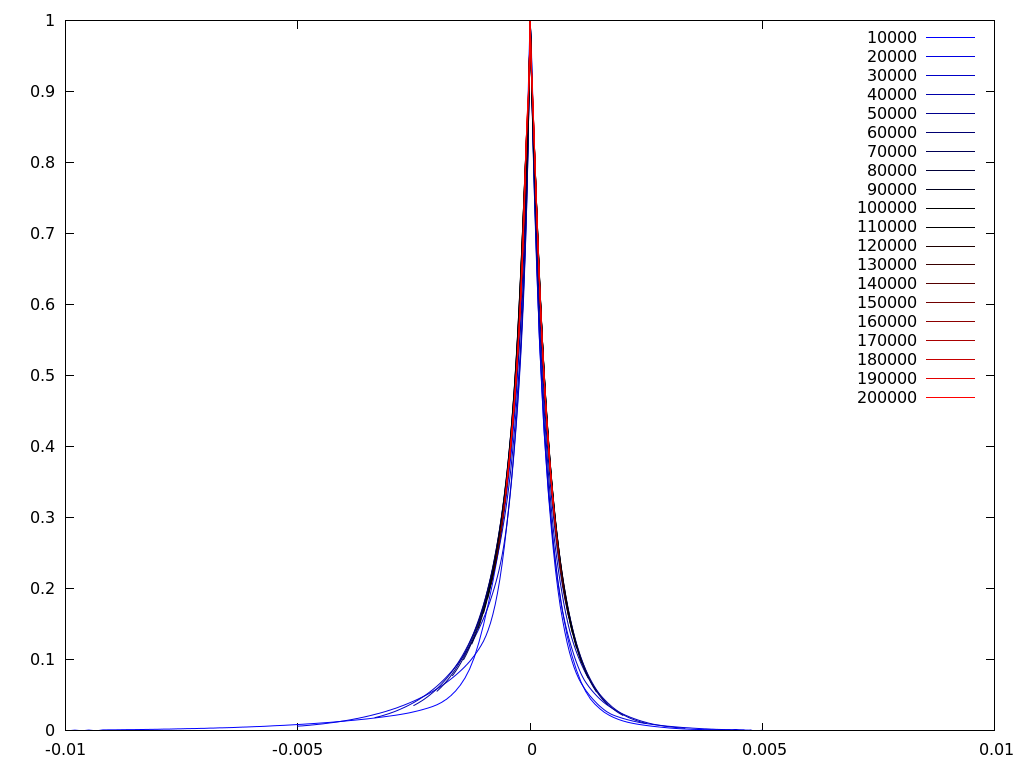}
\caption{The off-diagonal heat kernel $p_t(x,y)/p_t(x,x)$ plotted versus $|x-y|^\alpha/t$, where $x$ is the highest point from each of $6$ different realizations of the free field ($\gamma=0.8$, $\alpha=1$). The various times $t$, measured in the number of iterations of the Crank-Nicolson method, are indicated in the legend.}
\label{fig:08difffield}
\end{figure}

\begin{figure}[htp]
\centering
\includegraphics[scale=0.2]{alpha15-seed3-high03}
\includegraphics[scale=0.2]{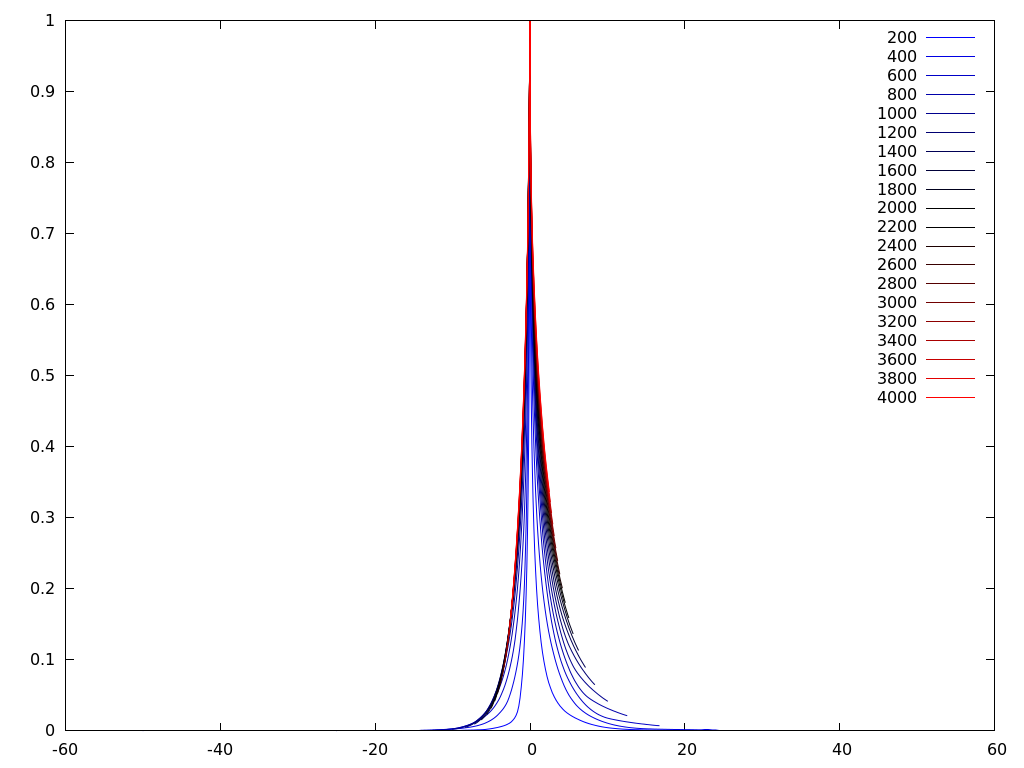}
\includegraphics[scale=0.2]{a1-seed5-high06}
\includegraphics[scale=0.2]{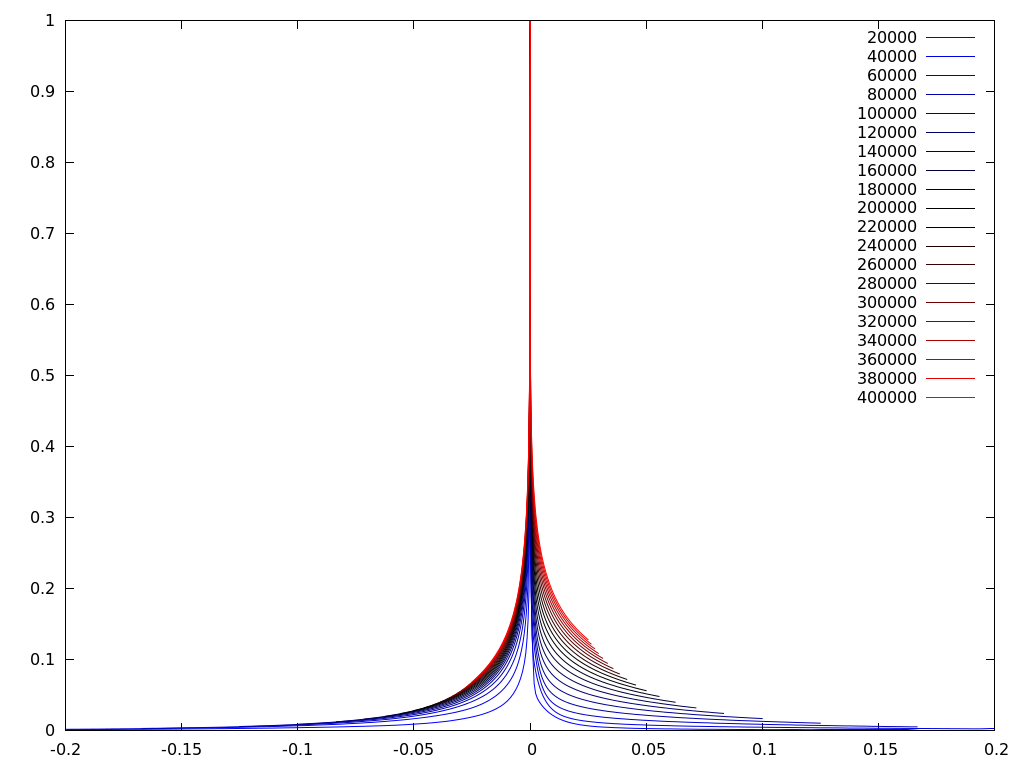}
\caption{The off-diagonal heat kernel $p_t(x,y)/p_t(x,x)$ plotted versus $|x-y|^\alpha/t$ for the same free field realization and the same starting (high) point, but fit to different values of $\alpha$ ($1$ on the left, $2$ on the right). In the top panels, a sample from $\gamma=0.4$. In the bottom panels, a sample from $\gamma=1.2$.}
\label{fig:diffalpha}
\end{figure}

For LQG models, we generate several realizations of the free field $X_n$, take $x$ to be each of the 10 highest points of $X_n$, and then compute the heat kernel for each pair $(X_n, x)$. Upon producing the heat ball snapshots as in Figure \ref{fig:heatballs}, we took a horizontal cut along the snapshot centered at $x$, that is, we extracted values of $p_t(x,y)$ for $y$ within a certain distance from $x$ and having the same horizontal coordinate as $x$. Given the rough spherical shape of the heat ball, we believe that it does not make a significant difference if one cuts the snapshot along a different direction for $y$. Then we plot $p_t(x,y)/p_t(x,x)$ versus $|x-y|^\alpha/t$. Some of these off-diagonal heat kernels are plotted in Figures \ref{fig:04} through \ref{fig:diffalpha}. Note that we are interested in the regime of \emph{small} $|x-y|^\alpha/t$.

Somewhat surprisingly, in an attempt to make the curves collapse onto one, we found that choosing $\alpha \in (1,2)$ produces the best fit; see Fig. \ref{fig:diffalpha}. (While not shown in the figure, we have found the fits worsen as $\alpha$ increases from $2$ onwards.) Moreover, as $\gamma$ increases toward $2$, the best-fit value of $\alpha$ decreases toward $1$. Although we do not resolve the value of $\alpha$ to high accuracy, the fits suggest that diffusion started from the highest points of the free field $X_n$ is \emph{super}-Gaussian ($\alpha<2$), rather than \emph{sub}-Gaussian ($\alpha>2$), as one often expects in the setting of fractal spaces. We provide some discussions of this curious finding in the next section.

More tests are needed to confirm whether super-Gaussian diffusion holds for large values of $|x-y|^\alpha/t$, as well as for \emph{generic} starting points on LQG. The results will be reported in a revised version of this report.

\section{Discussions}

In this section we would like to discuss our numerical findings in comparison with recent theoretical results on LQG.

\subsection{On-diagonal heat kernel and the spectral dimension}

It has been long conjectured in the physics literature that the spectral dimension of subcritical LQG (\emph{i.e.,} $\gamma\in [0,2)$) is $2$. From the mathematical point of view, one can introduce the local spectral dimension through the heat kernel as follows:
\begin{equation}
\label{eq:specdim}
d_s(x) = -2\lim_{t\downarrow 0} \frac{\log p_t(x,x)}{\log t},
\end{equation}
provided that this limit exists.

In \cite{RhodesVargasSpec} the authors proved a slightly weaker version of this statement, and essentially confirmed that $d_s(x)=2$ for $M_\gamma$-a.e. $x$.  Later, \cite{AndresKajino} strengthens the result by showing that both the upper and lower bounds on the on-diagonal heat kernel $p_t(x,x)$ is a constant multiple of $t^{-1}$ times a logarithmic function of $t^{-1}$, which implies by (\ref{eq:specdim}) that $d_s(x)$ is $2$ for $M_\gamma$-a.e. $x$.

Our numerics confirms the $t^{-1}$ dependence of the on-diagonal heat kernel. However due to discretization effects, we cannot, and do not, interpret the behavior for very short times $t$. For instance, since the $\log(t^{-1})$ dependence, if it indeed exists, becomes dominant only for very small $t$, we do not expect to see this effect from our numerics.

\subsection{Space-time scaling of the full heat kernel}

Next we would like to discuss aspects of space-time scaling in the heat kernel. The recent works of \cite{MRVZ} and \cite{AndresKajino}, respectively, established an off-diagonal upper bound on the heat kernel, as well as an on-diagonal (or near-diagonal) lower bound. To be precise, the authors of \cite{MRVZ} proved an off-diagonal upper bound and a uniform (non-matching) lower bound for the heat kernel on the unit torus $\mathbb{T}=([0,1]/\sim)^2$. (Below $\mathbb{P}^X$ stands for the Gaussian measure associated to the free field $X$.)

\textbf{Heat kernel upper bound on $\mathbb{T}$:} For all $\delta>0$, $\mathbb{P}^X$-a.s., there exists $\beta=\beta_\delta(\gamma)> 2$, and some positive random constants $c_1, c_2$ depending on $X$ only, such that
\begin{equation}
\label{eq:HKUB1}
p_t(x,y) \leq \left(\frac{c_1}{t^{1+\delta}}+1\right) \exp\left(-c_2 \left(\frac{|x-y|^\beta}{t}\right)^{1/(\beta-1)} \right).
\end{equation}
for all $x,y \in \mathbb{T}$ and all $t>0$.

\textbf{Heat kernel lower bound on $\mathbb{T}$:} Fix $x,y\in T$ and $\eta>0$. Then $\mathbb{P}^X$-a.s., there exists a random $T_0=T_0(X,x,y,\eta)>0$ such that for any $t\in [0,T_0]$,
\begin{equation}
p_t(x,y) \geq \exp\left(-t^{-\frac{1}{1+\gamma^2/4-\eta}}\right).
\end{equation}

Meanwhile, the authors of \cite{AndresKajino} proved an off-diagonal upper bound and an on-diagonal lower bound for the heat kernel on $\mathbb{R}^2$. The on-diagonal bounds differ by a logarithmic factor.

\textbf{Heat kernel upper bound on $\mathbb{R}^2$:} Fix a bounded domain $U\subset \mathbb{R}^2$. For any $\beta>\frac{1}{2}(\gamma+2)^2>2$, $\mathbb{P}^X$-a.s., there exist positive constants $c_i = c_i(X,\gamma,U,\beta)$ ($i=3,4$) such that
\begin{equation}
\label{eq:HKUBPf}
p_t(x,y) \leq c_3 t^{-1} \log(t^{-1}) \exp\left(-c_4\left(\frac{|x-y|^{\beta}}{t}\right)^{1/(\beta-1)}\right) 
\end{equation}
for all $t\in (0,\frac{1}{2}]$, $x\in \mathbb{R}^2$, and $y\in U$.

\textbf{Heat kernel lower bound on $\mathbb{R}^2$:} $\mathbb{P}^X$-a.s., for $M_{\gamma}$-a.e. $x\in\mathbb{R}^2$, there exist positive constants $c_5=c_5(X,\gamma)$ and $t_0=t_0(X,\gamma,x)>0$ such that
\begin{equation}
p_t(x,x) \geq c_5 t^{-1} (\log(t^{-1}))^{-\eta} \qquad \forall t\in (0,t_0],
\end{equation}
for an explicit positive constant $\eta$.

There is a well established theory for obtaining two-sided heat kernel bounds when the metric measure space is volume-doubling; see \emph{e.g.} \cite{GrigHuLau} and references therein. Given that the Liouville measure is probably not volume doubling (see Appendix A of \cite{BGRV}), it appears an open problem whether one can establish matching upper and lower heat kernel bounds.

Our numerical results on the heat kernel do not contradict the upper bound (\ref{eq:HKUB1}) or (\ref{eq:HKUBPf}). As \cites{MRVZ, AndresKajino} point out, the value of $\beta$ in the heat kernel upper bound may not be the best possible. However, perhaps the most puzzling finding from our numerical results is that diffusion on Liouville quantum gravity appears \emph{super}-diffusive with respect to the Euclidean metric ($\alpha<2$) for small values of $|x-y|^\alpha/t$, (at least) when started from the highest points of the free field. Given the authors' current knowledge, we cannot adequately explain this superdiffusive phenomenon, but we would like to offer some thoughts for inquiry.

First of all, whether the superdiffusivity with respect to the Euclidean metric is \emph{generic} for every (or just $M_{n,\gamma}$-a.e.) starting point remains to be investigated. Recall that on a two-dimensional domain, the Liouville measure $M_\gamma$ has full support on the set of $\gamma$-thick points (see \cite{HuMillerPeres} for definition) of the free field. The discrete analog of this notion is the $(\gamma/2)$-high point of the discrete free field described in \cite{Daviaud}. Based on this connection, it may be worth investigating on the lattice whether superdiffusivity persists when the starting point is chosen from other high points of the discrete free field. We do not rule out the possibility that such phenomenon may be non-generic, due to the dependence between the starting point of diffusion and the realization of the free field. (For example, the KPZ relation mentioned in \S1 holds for a generic set in LQG provided that the set is chosen independently of the free field. The KPZ relation may break down if the set chosen depends on the free field; see \cite{Aru} for the case where the sets are the level lines of the free field.)

On a more fundamental level, we suspect that this superdiffusive behavior is connected to the idea, long recognized by experts working in quantum gravity, that the Euclidean metric is \emph{not} the right metric to describe the geometry of LQG. It seems possible that by choosing a different metric $R(x,y)$ which is equivalent to some power of the Euclidean metric, say, $R(x,y) \asymp |x-y|^{2/\alpha}$, then diffusion becomes Gaussian with respect to this new metric. That being said, we do not claim that the $R$ metric has such a simple scaling relation with respect to the Euclidean metric. It is possible that due to the multifractal structure of the Liouville measure, there is a more complicated scaling relation between the putative geodesic metric and the Euclidean metric.

This brings us to the big open problem in this field, which is to identify the ``right'' random geodesic for LQG. One approach to finding the geodesic metric on a measure space is to use the intrinsic metric induced from the Dirichlet form for diffusion \cite{SturmI, SturmII, SturmIII, Stollmann}. But as pointed out in \cite{GarbanRhodesVargasII}*{\S4}, the intrinsic metric induced by the Dirichlet form for Liouville Brownian motion is trivial, and this has to do with the fact that the Liouville measure $M_\gamma$ is singular with respect to the Lebesgue measure (the ``energy measure'' of the Dirichlet form \cite{FOT}). In order to identify a nontrivial intrinsic metric, one would have to make a change of measure (equivalently, make a time change of the Liouville Brownian motion) so that the new measure is (minimally) energy dominant with respect to the Lebesgue measure \cites{Hino, HinzRocTep}. (For those coming from the analysis on fractals community, this problem is reminiscent of Brownian motion on the Sierpinski gasket, whose invariant measure is the self-similar measure $m$, which is singular with respect to the energy measure. In order to equip the gasket with a ``measurable Riemannian structure,'' one needs to change the measure to the so-called Kusuoka measure, which is minimally energy dominant in the sense of \cite{Hino}. See \cites{KigamiMRS} for details.) However, it is not clear to the authors what a ``canonical'' choice of the energy dominant measure should be on LQG.

At the time of this writing, there has been encouraging progress towards identifying the geodesic metric on LQG surface with $\gamma=\sqrt{8/3}$, see the program of Duplantier, Miller, and Sheffield as outlined in \cites{MSQLE, DMS}. We would like to ask: given that this metric $\mathfrak{d}$ is found, how does the heat kernel estimate depend on $\mathfrak{d}$? And does it lend to a measurable Riemannian structure on $(\mathbb{D}, M_{\gamma}, \mathfrak{d})$ so that one can study ``gradients'' (and not just Laplacians) of functions on LQG?

\subsection*{Acknowledgements}

We thank the UConn High Performance Computing (HPC) Center for allowing us to run our simulations on the HORNET cluster.

\vspace{20pt}

\begin{small}
\begin{flushleft}
\textbf{Grigory Bonik}, \textbf{Joe P. Chen} and \textbf{Alexander Teplyaev}\\
\textsc{Department of Mathematics, University of Connecticut, Storrs, CT 06269-3009, USA.}\\
\textsc{E-mails:} \href{mailto:gregory@bonik.org}{gregory@bonik.org}, \href{mailto:joe.p.chen@uconn.edu}{joe.p.chen@uconn.edu}, \href{mailto:alexander.teplyaev@uconn.edu}{alexander.teplyaev@uconn.edu}
\end{flushleft}
\end{small}

\end{document}